\documentclass{jfm}
\usepackage{float,graphicx,psfrag,subfigure,enumerate}
\usepackage{setspace}
\usepackage{bm}
\usepackage{mathtools}
\usepackage{color}
\usepackage{lineno}
\usepackage{cases}
\usepackage{booktabs}

\definecolor{dgreen}{rgb}{0.0,0.45,0.0}

\newcommand{\rone}[1]{{\color{black} #1}}
\newcommand{\rtwo}[1]{{\color{black} #1}}
\newcommand{\rthree}[1]{{\color{black} #1}}

\graphicspath{{figures/}}

\shorttitle{Modelling the turbulent burning velocity using propagating surfaces}
\shortauthor{J. You and Y. Yang}

\title{Modelling of the turbulent burning velocity based on Lagrangian statistics of propagating surfaces}

\author{Jiaping You\aff{1} and Yue Yang\aff{1,2} \corresp{\email{yyg@pku.edu.cn}}}

\affiliation{\aff{1}State Key Laboratory for Turbulence and Complex Systems, College of Engineering,\\ Peking University, Beijing 100871, China
\aff{2}CAPT and BIC-ESAT, Peking University, Beijing 100871, China}

\newcommand{\bs}{\boldsymbol}

\newcommand{\mc}{\mathcal}
\newcommand{\rd}{\textrm{d}}

\newcommand{\EQ}{\begin{equation}}
\newcommand{\EN}{\end{equation}}
\newcommand{\EQA}{\begin{eqnarray}}
\newcommand{\ENA}{\end{eqnarray}}
\newcommand{\D}[2]{\frac{\partial #1}{\partial #2}}

\newcommand{\de}[2]{\frac{\textrm{d} #1}{\textrm{d} #2}}

\newcommand{\<}{\langle}
\renewcommand{\>}{\rangle}

\newcommand{\Rmnum}[1]{\expandafter\@slowromancap\romannumeral #1@}

\begin{document}

\maketitle
\begin{abstract}

We propose a predictive model of the turbulent burning velocity $S_T$ \rtwo{in homogeneous isotropic turbulence (HIT)} based on Lagrangian statistics of propagating surfaces.
The propagating surfaces with a constant displacement speed are initially arranged on a plane, and they evolve in non-reacting HIT, behaving like the propagation of a planar premixed flame front.
%
The universal constants in the model of $S_T$ characterize the enhancement of area growth of premixed flames by turbulence, and they are determined by Lagrangian statistics of propagating surfaces.
The flame area is then modelled by the area of propagating surfaces at a truncation time.
This truncation time signals the statistical stationary state of the evolutionary geometry of propagating surfaces, and it is modelled by an explicit expression using limiting conditions of very weak and strong turbulence.
Another parameter in the model of $S_T$ characterizes the effect of fuel chemistry on $S_T$, and it is pre-determined by very few available data points of $S_T$ from experiments or direct numerical simulation (DNS) in weak turbulence.
The proposed model is validated using three DNS series of turbulent premixed flames with various fuels. The model prediction of $S_T$ generally agrees well with DNS in a wide range of premixed combustion regimes, and it captures the basic trends of $S_T$ in terms of the turbulence intensity, including the linear growth in weak turbulence and the `bending effect' in strong turbulence.
%

\end{abstract}


\section{Introduction}\label{sec:intro}
The turbulent burning velocity $S_T$ measures how fast a premixed flame consumes the fresh fuel mixture in turbulence, and characterizes the impact of turbulence on the enhancement of \rtwo{burning rate}.
The determination of $S_T$ is one of the most important unsolved problems in turbulent premixed combustion \citep[]{Peters2000}, and is critical in many models of turbulent premixed flames \citep[][]{Lipatnikov2002,Driscoll2008}.

In general, $S_T$ depends on the turbulence intensity $u'$, flame geometry, fuel chemistry \citep[see][]{Lipatnikov2002,Driscoll2008} and gas expansion \citep[][]{Peters2000,Sabelnikov2017}.
%
Recent studies also investigate the dependence of $S_T$ on the low temperature fuel chemistry \citep[][]{Won2014}, high pressure \citep[][]{Bradley2013, Venkateswaran2014}, high turbulence intensity \rtwo{\citep[][]{Wabel2017,Nivarti2019}}, flame stretch rates and curvatures \citep[][]{Thiesset2017}.

Although there are a variety of empirical models \citep[\eg][]{Bradley1992,Lipatnikov2002,Driscoll2008} for predicting $S_T$ and the related flame area ratio, the predictions from different models cannot collapse owing to the lack of a rigourous theoretical framework and a unique definition for $S_T$, and to the uncertainties of different measurement methods and flame geometries.
Various power laws are used to fit data points from the experimental measurement of $S_T$ in terms of $u'$, but most expressions depend on empirical parameters \citep[\eg][]{Bradley1992,Peters1999,Lipatnikov2002,Driscoll2008,Venkateswaran2014,Wabel2017}, and the corresponding scaling exponent is sensitive to flame configurations and measurements \citep[see][]{Verma2016}.
Therefore, a consensus on the universal model for predicting $S_T$ versus $u'$ from experimental data remains elusive.
In addition, it is possible to use the direct numerical simulation (DNS) to calculate $S_T$ for turbulent premixed flames with the simple geometry and moderate turbulent Reynolds number $Re$ \rthree{\citep[\eg][]{Tanahashi2000,Bell2005,Bell2013,Wang2017JFM}}, but the computational cost is formidable for engineering applications.

In a theoretical framework, $S_T$ can be modelled by the Eulerian or the Lagrangian approach. Both approaches generally presume that the premixed flame is in the flamelet regime, and this assumption can be extended to the corrugated flame and thin-reaction-zone regimes with further modelling efforts \citep[\eg][]{Peters1999}.
%

In the Eulerian approach, the local geometry and topology of the flame surface is essential to modelling methods, e.g.~the flame surface density (FSD) \citep[see][]{Pope1988,Candel1990,Trouve1994,Veynante2002}, G-equation model \citep[see][]{Peters2000}, level-set methods \citep[\eg][]{Creta2011,Yu2015} and fractal models \citep[\eg][]{Gouldin1987,Knikker2002,Fuerby2005}.
In general, the flame front is extracted as the isosurface of an auxiliary scalar function, and the scalar is evolved through a partial differential equation.
The additional transport equation of the Reynolds-averaged or filtered scalar field needs to be solved along with modelled conservation equations, and the modelled equations involve some empirical constants and unclosed terms which have to be modelled by further assumptions.
The Eulerian modelling methods are usually used in the context of the large-eddy simulation and Reynolds-averaged Navier--Stokes simulation.
Although their computational cost can be much lower than that of combustion DNS, most of them are not able or not intended to be converted to explicit expressions of $S_T$ as those obtained from experiments in terms of integral quantities.

Compared with the Eulerian approach, the Lagrangian approach appears to be more natural to describe flame wrinkling with a `memory' of any wrinkling occurring upstream \citep[see][]{Driscoll2008,Zhou2019}, but the Lagrangian modelling is generally restricted to non-reacting turbulence.
\citet{Pope1988} established a rigorous formulation to describe the Lagrangian evolution of surface elements in turbulent flows.
Up to date, the Lagrangian approach has been extensively applied to investigate the local dynamics of non-reacting turbulence using material or propagating surface elements \citep[][]{Girimaji1990,Girimaji1992,Zheng2017}, Lagrangian particles \citep[see][]{Yeung2002,Toschi2009} and Lagrangian scalars \citep[][]{Yang2010a}.
In particular, \citet{Girimaji1992} investigated Lagrangian statistics of the tangential strain rate and characteristic curvature of propagating surface elements in homogeneous isotropic turbulence (HIT), which helps to understand the deformation of premixed flames under turbulent straining motion.

Recently, the Lagrangian approach also emerges as a diagnostic tool for investigating turbulence--flame interactions in turbulent combustion.
\citet{Steinberg2015} quantified variations of the vorticity and strain rate of fluid parcels undergoing combustion by tracking Lagrangian fluid particles and using the three-dimensional, time-resolved experimental measurement in premixed dimethyl-ether/air piloted jet flames.
\citet{Day2015} observed the complex nature of thermodynamic and chemical evolutions along Lagrangian trajectories, including the non-monotonic evolution of temperature within fluid parcels, in a joint experimental and computational study of lean hydrogen turbulent premixed flames in low swirl burners.
\citet{Hamlington2017} examined the effects of high-speed turbulence on the non-monotonic thermochemical trajectories in DNS of hydrogen-air premixed flames.
\citet{Chaudhuri2015} developed a method of Lagrangian flame particles to study the time-history of specific portions and associated properties of the flame surface. This method has been utilized to analyze turbulence--chemistry interactions \citep[][]{Chaudhuri2015} and extinction dynamics in H$_2$/air premixed flames \citep[][]{Uranakara2016}. \cite{Dave2018} developed a backward flame-particle tracking method to locate the origin of the complex topology and physico-chemical state in a fully developed turbulent premixed flame.

Although the Lagrangian investigation on turbulent premixed flames serves as a valuable diagnostic tool for gaining physical insights by interrogating combustion DNS data, there is a lack of predictive models of $S_T$ from the Lagrangian perspective.
Thus we aim to incorporate the statistical information of Lagrangian surface elements into the modelling of $S_T$, which can bridge the gap between the studies of Lagrangian statistics in non-reacting turbulence and Lagrangian-based diagnostics in turbulent premixed combustion.

The present modelling framework of $S_T$ is based on Lagrangian statistics of an ensemble of propagating surface elements.
The propagating surface elements, which approximate the flame surface, evolve independently via a set of ordinary differential equations for Lagrangian quantities \citep[see][]{Pope1988,Girimaji1992,Zheng2017}.
Each surface element is driven by a local fluid velocity and a displacement velocity normal to itself.
The influence of molecular diffusion and chemical reaction on the motion of the flame front is only through the displacement velocity.
As a simple model, the propagating surface decouples the diffusion and chemical reaction from the hydrodynamics effect on turbulent premixed combustion.

In the present study, we explore connections between $S_T$ and statistical geometry of propagating surfaces from a Lagrangian viewpoint, and then propose a simple predictive model of $S_T$ in terms of $u'$.
Most of the model parameters are pre-determined by Lagrangian statistics of propagating surfaces during a short time period in non-reacting HIT. 
%
%
We remark that the modelling of $S_T$ can be very sensitive to the flame geometry, \rtwo{so the present flow configuration and the applicability of the proposed $S_T$ model are restricted to HIT}.

The outline of this paper is as follows. In \S\,\ref{sec:numer}, we describe the numerical methods for the combustion DNS and tracking of propagating surfaces. In \S\,\ref{sec:mdoeldevelop}, we discuss the area growth of propagating surfaces, and then develop a predictive model of $S_T$ based on the Lagrangian statistics and several physical assumptions. In \S\,\ref{sec:assessST},  we present both \textit{a posteriori} and \textit{a priori} tests for the proposed model. Some conclusions are drawn in \S\,\ref{sec:conclusions}.

\section{Numerical overview}\label{sec:numer}

\subsection{Combustion DNS}\label{sec:CDNS}

\begin{figure}
  \centering
  \includegraphics[width=0.9\textwidth]{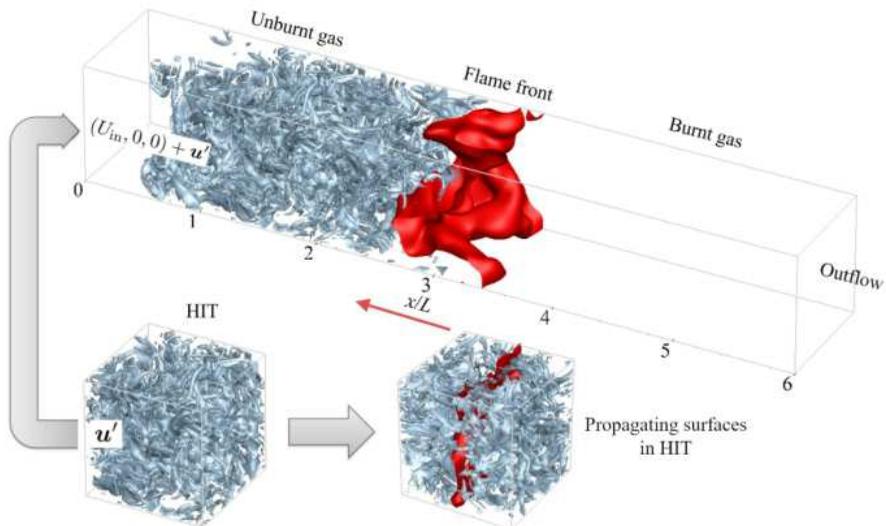}
  \caption{(Colour online)~A schematic diagram of DNS configurations of the premixed flame propagation (top) and the non-reacting HIT (bottom), with light blue isosurfaces of $|\bs \omega|=3\times 10^5$ s$^{-1}$, the red flame front (top) and the red global propagating surface (bottom). The non-reacting HIT is utilized for the inflow condition in combustion DNS and for the tracking of propagating surfaces. }
  \label{fig:config}
\end{figure}

For the combustion DNS, we consider the free propagation of a planar H$_2$/air premixed flame along the streamwise $x$-direction in statistically stationary HIT.
\rthree{This inflow-outflow DNS configuration of turbulent flames \citep[\eg][]{Tanahashi2000,Aspden2011,Savard2015,Nivarti2017,Minamoto2018}} is presented in figure~\ref{fig:config}. The computational domain is a cuboid with sides $L_x \times L_y \times L_z=6L \times L \times L$ and $L=2$ mm. It has inflow and outflow conditions at left and right boundaries, respectively, and periodic boundary conditions in lateral $y$- and $z$-directions. This domain is discretized on uniform grid points $N_x \times N_y \times N_z=6N \times N \times N$ \rthree{with the mesh spacing $\Delta x$}.

The unburnt gas is a lean H$_2$/air mixture with the equivalence ratio $\phi=0.6$ at the temperature $T_{u}=300$ K and atmospheric pressure.
It is ignited by a planar laminar flame initially located in the middle of the computational domain.
The one-dimensional steady laminar premixed flame is computed using the PREMIX code \citep[][]{Kee85} with the nine-species detailed kinetic mechanism \citep[][]{Li2004} and mixture-averaged transport properties. The laminar flame speed is $S_L=0.723$ m/s, the flame thermal thickness \rthree{ $\delta_L \equiv (T_b-T_u)/|\nabla T|_{\max}=0.361$ mm, and the flame time scale $\tau_f\equiv\delta_L/S_L=0.499$ ms, where $T_u$ and $T_b$ are temperatures in unburnt and burnt mixtures, respectively, and $|\nabla T|_{\max}$ denotes the maximum temperature gradient in the laminar premixed flame.}

We carried out five DNS cases with varied turbulent intensity $u'$ listed in table~\ref{tab:params}.
\rthree{For the inflow condition, we first performed a separate DNS of non-reacting, statistically stationary HIT. This HIT data is then imposed to the bulk inflow velocity $U_{\textrm{in}}(t)$ by using Taylor's hypothesis.
Thus the inflow condition is $\bs {u}_{\textrm{in}}(y,z,t)=U_{\textrm{in}}(t)+\bs{u}'(y,z,t)$ with the fluctuating velocity $\bs{u}'(y,z,t)$ on the $y$--$z$ plane moving through the HIT field at $U_{\textrm{in}}(t)$.}
A feedback control algorithm \citep[][]{Bell2006} is applied to dynamically adjust $U_{\text{in}}(t)$ to stabilize the turbulent flame in the middle of the computational domain. \rthree{The convective condition \citep[][]{Desjardins08} is used for the velocity and scalars at the outflow boundary in the streamwise direction.}
A stable, linear velocity forcing \citep[][]{Carroll2013,Savard2015} is adopted to maintain the turbulent intensity from $x=0.5L$ to $5L$ along the streamwise direction in combustion DNS and in the entire domain in non-reacting DNS.

\begin{figure}
  \centering
  \includegraphics[width=0.6\textwidth]{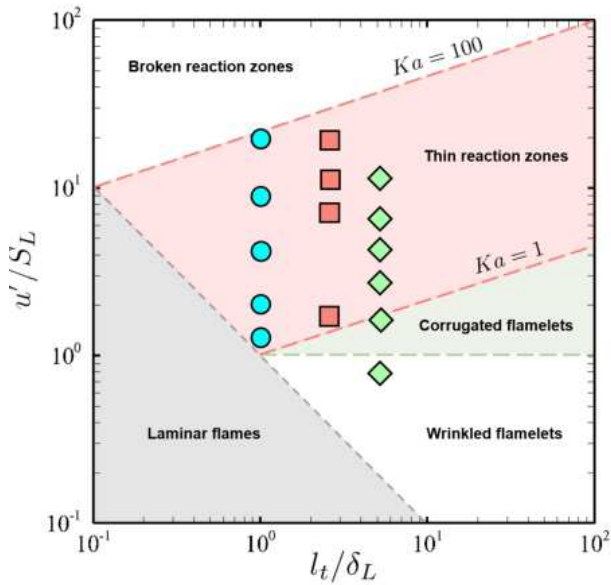}
  \caption{(Colour online)~Parameters of three DNS series in the regime diagram of turbulent premixed combustion. Circles: the present DNS; squares: DNS of \citet{Nivarti2017}; diamonds: DNS (group T) of \citet{Lee2010}.}
  \label{fig:regime}
\end{figure}

The integral length scale $l_t$, the eddy turnover time $T_e$, the Kolmogorov time scale $\tau_\eta$ and length scale $\eta$, and the turbulent Reynolds number $Re \equiv u'l_t/\nu$ of the HIT are summarized in table~\ref{tab:params}. The Karlovitz number $Ka\equiv\left(u'/S_L\right)^{3/2}\left(\delta_L/l_t\right)^{1/2}$  and the Damk\"{o}hler number $Da\equiv\left(S_L/u'\right)\left(l_t/\delta_L\right)$ are computed in the unburnt side.
In terms of the regime diagram \citep[][]{Peters2000} in figure~\ref{fig:regime}, case A is very close to the regime of wrinkled flamelets and laminar flames, cases B, C and D are in the thin reaction zone and case E is close to the broken reaction zone.
\rthree{We remark that the boundaries dividing different regimes in the Borghi diagram in figure~\ref{fig:regime} are solely based on phenomenological arguments and dimensional analysis. Thus the applicability of the diagram is still under debate for accurately characterizing flame structures \citep[\eg][]{Aspden2011,Tamadonfar2015,Wabel2017,Aspden2017,Skiba2018}, and the diagram is only presented for roughly comparing flow and flame parameters in different DNS series.}

The numerical resolution in all the cases is ensured to resolve the smallest turbulent scales by \rthree{the criterion $\eta/\Delta x>0.5$} \citep[see][]{Pope2000} and to resolve the flame length scales using a minimum of 24 grid points within $\delta_L$. 
\rthree{In addition, our grid convergence study (not shown) further validates that the present combustion DNS is sufficiently well-resolved by comparing $S_T$ and flame structures from simulations on the present grid and a coarser grid.}

%


\begin{table}
  \centering
  \setlength{\tabcolsep}{3.5mm}{
  \begin{tabular}{ccccccc}

  Case &A   &B  &C  &D  &E  \\

  $N $                     &128      &128      &128      &\rthree{256}       &256     \\
  $u'/S_L$                 &1.0      &2.0      &5.0      &10.0      &20.0    \\
  $l_t/\delta_L$           &0.91     &0.91     &0.91     &0.91      &0.91    \\
  $T_e$ ($\mu$s)           &1100     &426      &156      &72.7      &38.3    \\
  $\tau_\eta$ ($\mu$s)     &144      &47.6     &11.9     &4.21      &1.53    \\
  $\eta $ ($\mu$m)         &52.9     &30.4     &15.2     &9.04      &5.45    \\
  \rthree{$\eta/\Delta x$} &\rthree{3.39}     &\rthree{1.95}     &\rthree{0.97}     &\rthree{1.16}      &\rthree{0.70}     \\
  \rthree{$\Rey$}          &\rthree{14.24}    &\rthree{28.48}    &\rthree{71.2}     &\rthree{142.4}     &\rthree{284.8}    \\
  $\mbox{\textit{Ka}}$     &1.04     &2.94     &11.64    &32.91     &93.08    \\
  $\mbox{\textit{Da}}$     &0.92     &0.46     &0.18     &0.09      &0.05     \\

\end{tabular}}
\caption{Parameters for DNS cases.}
\label{tab:params}
\end{table}

The low-Mach-number, variable-density formulation of conservation equations of mass, momentum, species and energy are solved on staggered grid points using the NGA code \citep[][]{Desjardins08}.
A second-order centered, kinetic-energy conservative finite difference scheme is used for discretizing spatial derivatives in momentum equations. \rthree{A third-order bounded QUICK scheme \citep[][]{Herrmann2006} with low numerical dissipation and preserved physical bounds of scalars is adopted for treating convection terms in scalar transport equations of species mass fractions and temperature, and this scheme has been used in a series of DNS of turbulent premixed flames \citep[\eg][]{Savard2015,Bobbitt2016}.} The temporal integration of the conservation equations is advanced by an iterative semi-implicit Crank--Nicolson scheme \citep[][]{Pierce01} with the Strang splitting between transport and chemistry.
The detailed nine-species H$_2$/air kinetic mechanism \citep[][]{Li2004} is employed with reaction rates, thermodynamic properties evaluated by the CHEMKIN \citep[][]{Kee96} library.
The constant Lewis numbers listed in table~\ref{tab:Lenumbers} are used in DNS and they are determined from the fit in terms of mixture-averaged transport properties for the laminar premixed flame.
The time integration of chemical source terms is performed using the stiff DVODE solver \citep[][]{Brown1989}.
Each DNS case is first run for at least $10T_e$ to reach a statistically stationary state, and then statistical properties of interest are calculated over a period of $20T_e$.

\begin{table}
  \centering
  \setlength{\tabcolsep}{2.5mm}{
  \begin{tabular}{lccccccccccccccc}
  &Species   &H$_2$    &O$_2$    &H$_2$O   &H$_2$O$_2$   &HO$_2$   &OH     &H      &O      &N$_2$   \\
  &$Le$      &0.27      &1.04      &0.79      &1.03      &1.04     &0.67   &0.16   &0.66   &1.01    \\
  \end{tabular}}
  \caption {Constant species Lewis numbers.}
  \label{tab:Lenumbers}
\end{table}

We define a reaction progress variable
\EQ
c\equiv \frac{Y_{\textrm{f}}-Y_{\textrm{f,u}}} {Y_{\textrm{f,b}}-Y_{\textrm{f,u}} }
\label{eq:progdef}
\EN
to characterize the progress of reaction and flame propagation, where $Y_{\text{f}}$ is the fuel mass fraction,  $Y_{\textrm{f,u}}$ and $Y_{\textrm{f,b}}$ are the fuel mass fraction in unburnt and burnt mixtures, respectively. The isosurface of $c=\hat{c}$ propagates at a displacement speed $S_d$.
The instantaneous flame front is chosen such that $\hat{c}=0.8$ corresponds to the location of the maximum heat release rate in the unstrained laminar flame.

\begin{figure}
  \centering

  \includegraphics[width=0.8\textwidth]{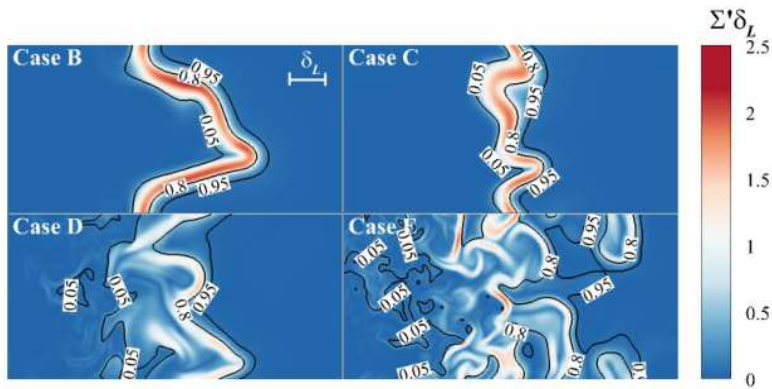}
  \caption{(Colour online)~Normalized local FSD  $\Sigma'\delta_L$ on a slice of sizes $2L \times L$ centered at $x=3L$ and at $t=20T_e$ from four combustion DNS cases.} 
  \label{fig:FSD}
\end{figure}

Figure~\ref{fig:FSD} depicts contours of the normalized local FSD $\Sigma'\delta_L$ on a slice near the flame front at $t=20T_e$ in combustion DNS cases B, C, D and E, with three isolines of $c=0.05, 0.8$ and $0.95$ for characterizing the turbulent flame brush.
Here, the large local FSD generally characterizes the flame front, and the detailed definition of FSD is explained in Appendix~\ref{sec:FSD}.
The laminar flamelet structure is retained in the most part of the flame brush in case B, and the flamelet structure in the preheat zone is disturbed and progressively broadened in cases C and D.
The topology of the reaction layer, as the isoline of $c=0.8$, still retains sheet-like parallel surfaces in both cases in the thin reaction zone. With increasing $Ka$, some portions of the flame front are broken down in case E.

\subsection{Tracking of propagating surface elements} \label{sec:track}

As sketched in figure~\ref{fig:config}, the non-reacting HIT field used in the inflow condition of combustion DNS is also utilized in the tracking of propagating surface elements.
For each DNS case listed in table~\ref{tab:params}, an ensemble of infinitesimal propagating surface elements are initially arranged on a planar surface in non-reacting HIT to model the propagation of a turbulent premixed flame.

Each propagating surface element carries the information of its Lagrangian location $\bs{X}(t)$, local coordinate system $\bs e_i(\bs{X}(t),t)$, curvature tensor $h_{\alpha\beta}$, and nominal surface area $\delta A$ \citep[][]{Pope1988}.
The evolution equation of $\bs{X}(t)$ is
 \EQ
 \de{\bs{X}(t)}{t}=\bs{u}(\bs{X}(t),t)+S_d(\bs{X}(t),t)\bs{n}(\bs{X}(t),t).
 \label{eq:loc}
 \EN
Each surface element moves with the local fluid velocity $\bs u$ and the displacement velocity $S_d \bs n$, where $\bs n$ denotes the unit normal of the surface element, and the displacement speed $S_d$ can be simply modelled by a constant $S_L$ or a variable accounting for the effect of weak flame stretch.

\begin{figure}
  \centering
  \includegraphics[width=0.98\textwidth]{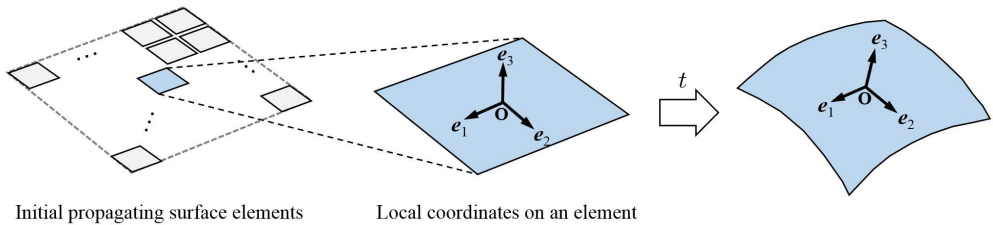}
  \caption{(Colour online)~Sketch of the temporal evolution of propagating surface elements.}
  \label{fig:surfaceelement}
\end{figure}

As sketched in figure~\ref{fig:surfaceelement}, a local Cartesian coordinate system $\bs e_i(\bs{X}(t),t)$, $i=1,2,3$ is attached to each surface element in order to compute $\bs n$ and other geometric properties of surface elements. Its origin denoted by `O' is at the position $\bs X(t)$ of the surface element. The unit normal vector $\bs{e}_3$ is equivalent to $\bs n$ in (\ref{eq:loc}), and two orthogonal unit vectors $\bs{e}_1$ and $\bs{e}_2$ span the tangent plane.

The governing equations for this coordinate system are
 \EQA
 \de{\bs{e}_3}{t} &=& -\bs{e}_{\alpha}u_{3,\alpha}, \nonumber \\
 \de{\bs{e}_{\alpha}}{t} &=& \frac{1}{2}\bs{e}_{\beta}(u_{\beta,\alpha}-u_{\alpha,\beta})+\bs{e}_3 u_{3,\alpha},~~~\alpha, \beta=1, 2,
 \label{eq:axis}
 \ENA
where `$,\alpha$' denotes the partial derivative in the direction of $\bs{e}_{\alpha}$.
The governing equation of the curvature tensor $h_{\alpha\beta}$ is 
\EQ
\de{h_{\alpha\beta}}{t}=s_{33}h_{\alpha\beta}-(s_{\gamma\beta}h_{\alpha\gamma}+s_{\gamma\alpha}h_{\beta\gamma})
    +u_{3,\alpha\beta}+S_d h_{\alpha\gamma}h_{\gamma\beta},
\label{eq:curve}
\EN
where the partial derivatives of $S_d$ are neglected \citep[][]{Zheng2017} and
\EQ
s_{ij} \equiv \frac{1}{2} \left(\frac{\partial u_i} {\partial x_j}+ \frac{\partial u_j}{\partial x_i }\right),~~~i,j=1,2,3
\label{eq:sr_tensor}
\EN
denotes the local rate-of-strain tensor.
Principal curvatures $\kappa_i$, $i=1,2$ of a surface element are defined as $\kappa_i = -k_i$ with $\kappa_1>\kappa_2$, where $k_i$, $i=1,2$ are eigenvalues of $h_{\alpha\beta}$.
The surface element convex (or concave) to the propagating direction $\bs n$ has a positive (or negative) curvature.

At time $t$, $\delta A(t)$ denotes the nominal surface area which can also be considered as the area ratio of infinitesimal surface elements.
The governing equation of $\delta A(t)$ or the surface stretch rate $K$ of the propagating surface is
\EQ
K \equiv \frac{1}{\delta A(t)} \de{\delta A(t)}{t} = K_t + 2{S_d}\kappa,
\label{eq:area}
\EN
where the tangential strain rate $K_t\equiv u_{1,1}+u_{2,2}$ characterizes the stretching of the surface area due to straining by the flow field, and $2{S_d}\kappa$ with the mean curvature $\kappa \equiv(\kappa_1+\kappa_2)/2$ represents the factional rate of change of area due to propagation.

\begin{figure}
  \centering
  \psfrag{a}[c][c]{\footnotesize $t/t_\eta$}
  \psfrag{b}[c][c]{\footnotesize $A(t)/A_0$}
  \psfrag{h}{\footnotesize $N_e=128^2$}
  \psfrag{i}{\footnotesize $N_e=256^2$}
  \psfrag{j}{\footnotesize $N_e=512^2$}

  \includegraphics[width=0.5\textwidth]{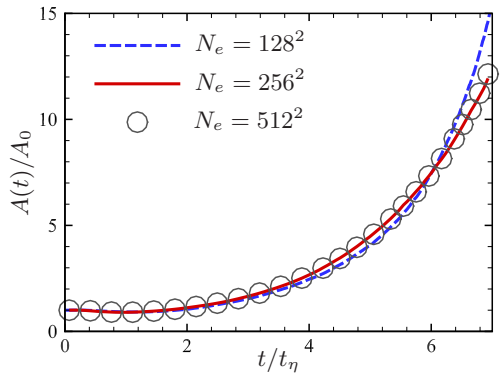}
  \caption{(Colour online)~Convergence test of the number of propagating surface elements in non-reacting DNS case C.}
  \label{fig:PartticleNumber}
\end{figure}

At the initial time, $N_e=256^2$ surface elements are uniformly arranged on a $x$--$y$ plane at $z=L/2$ to model a propagating planar flame in premixed combustion (see figures~\ref{fig:config} and \ref{fig:surfaceelement}).
Each surface element has $(\bs e_1,\bs e_2,\bs e_3)=(0,0,1)$, $h_{\alpha\beta}=0$, and $\delta A(0)=A_0/N_e$, where $A_0 \equiv \sum\limits_{N_e}\delta A(0)$ is the initial global surface area.
A convergence test with increasing $N_e$ in figure~\ref{fig:PartticleNumber} shows that the total surface area ratio, which is defined later in \S\,\ref{sec:areagrowth}, from $N_e=256^2$ surface elements are almost identical to those from $N_e=512^2$, so $N_e=256^2$ is used in the present study.

The tracking of propagating surface elements begins when the non-reacting HIT has reached the statistically stationary state.
A second-order, symmetry-preserved Runge--Kutta scheme \citep[][]{Zheng2017} is applied to advance (\ref{eq:loc}), (\ref{eq:axis}) and (\ref{eq:area}) explicitly and (\ref{eq:curve}) implicitly.
The time increment is sufficiently small to resolve the finest scales of the velocity field.
The Fourier spectral method is used to evaluate the velocity derivatives involved in these equations.
The interpolation is performed by a sixth-order Lagrangian interpolation scheme.
In addition, the propagating surface elements can evolve into cusps after a finite time \citep[][]{Girimaji1992}, and these cusps are removed in the tracking based on the positive definiteness of the curvature tensor.
The detailed numerical implementation of the surface tracking and the criterion for detecting cusps are described in \citet{Zheng2017}, and numerical errors have been verified to be negligible.
It is noted that the computational cost for the tracking of propagating surfaces in non-reacting DNS is very low as $O(1)$ CPU hours on a single core.
By contrast, the computational cost of each combustion DNS on a parallel machine is $O(10^4)\sim O(10^5)$ CPU hours.

\begin{figure}
 \centering
 \subfigure{
 \centering
 \includegraphics[width=0.4\textwidth]{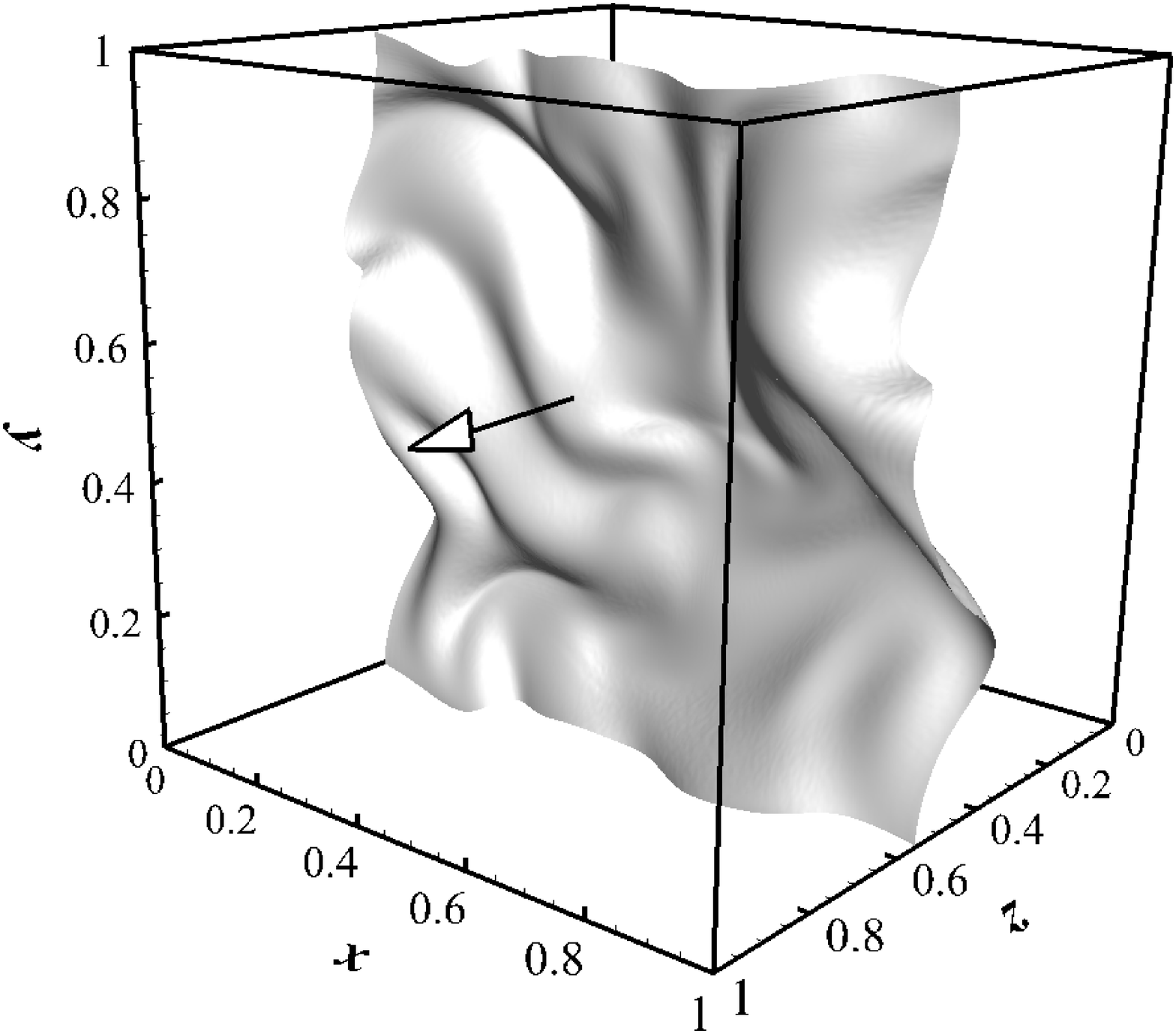}}
 \subfigure{
 \centering
 \includegraphics[width=0.4\textwidth]{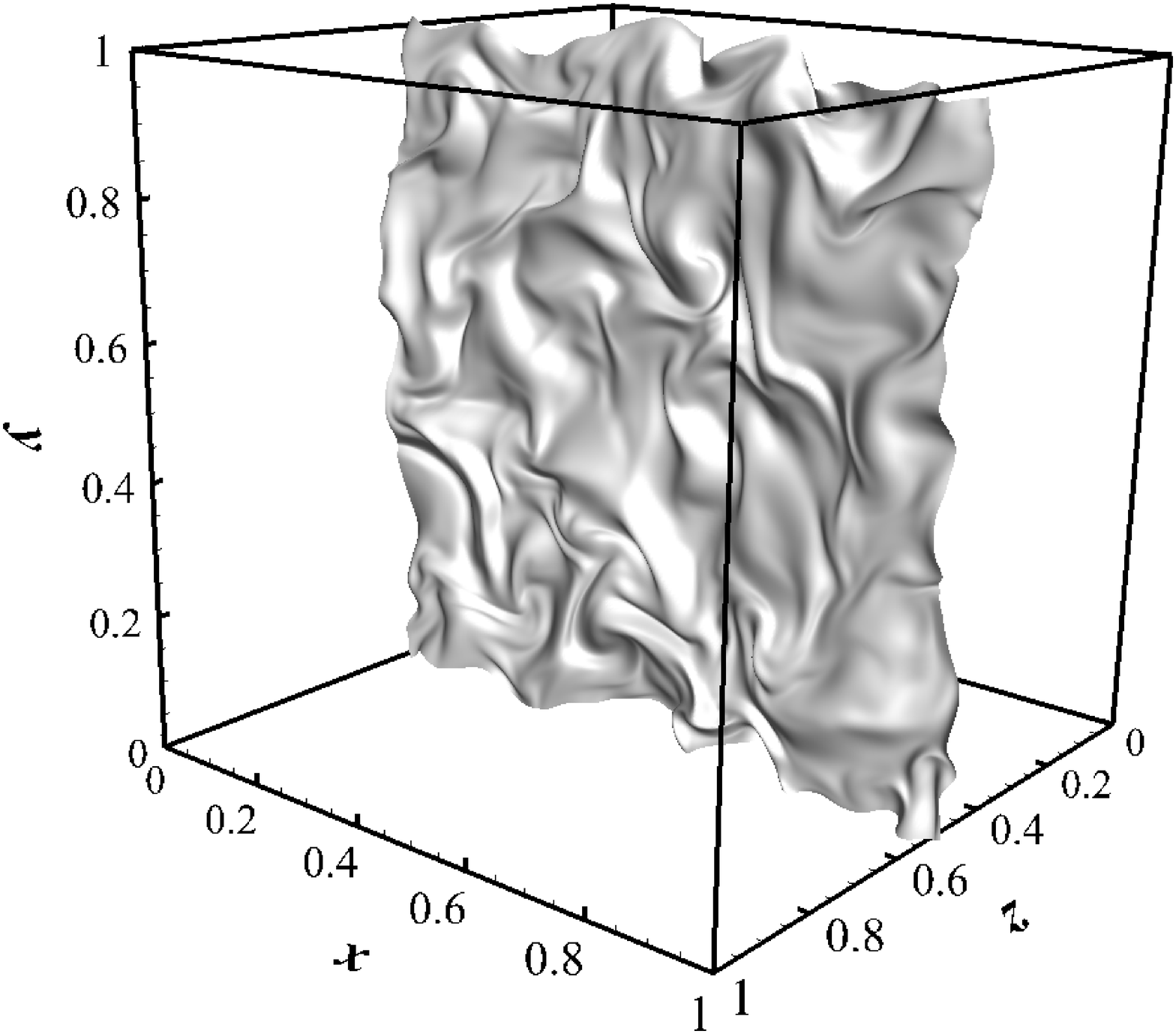}}

 \subfigure{
 \centering
 \includegraphics[width=0.4\textwidth]{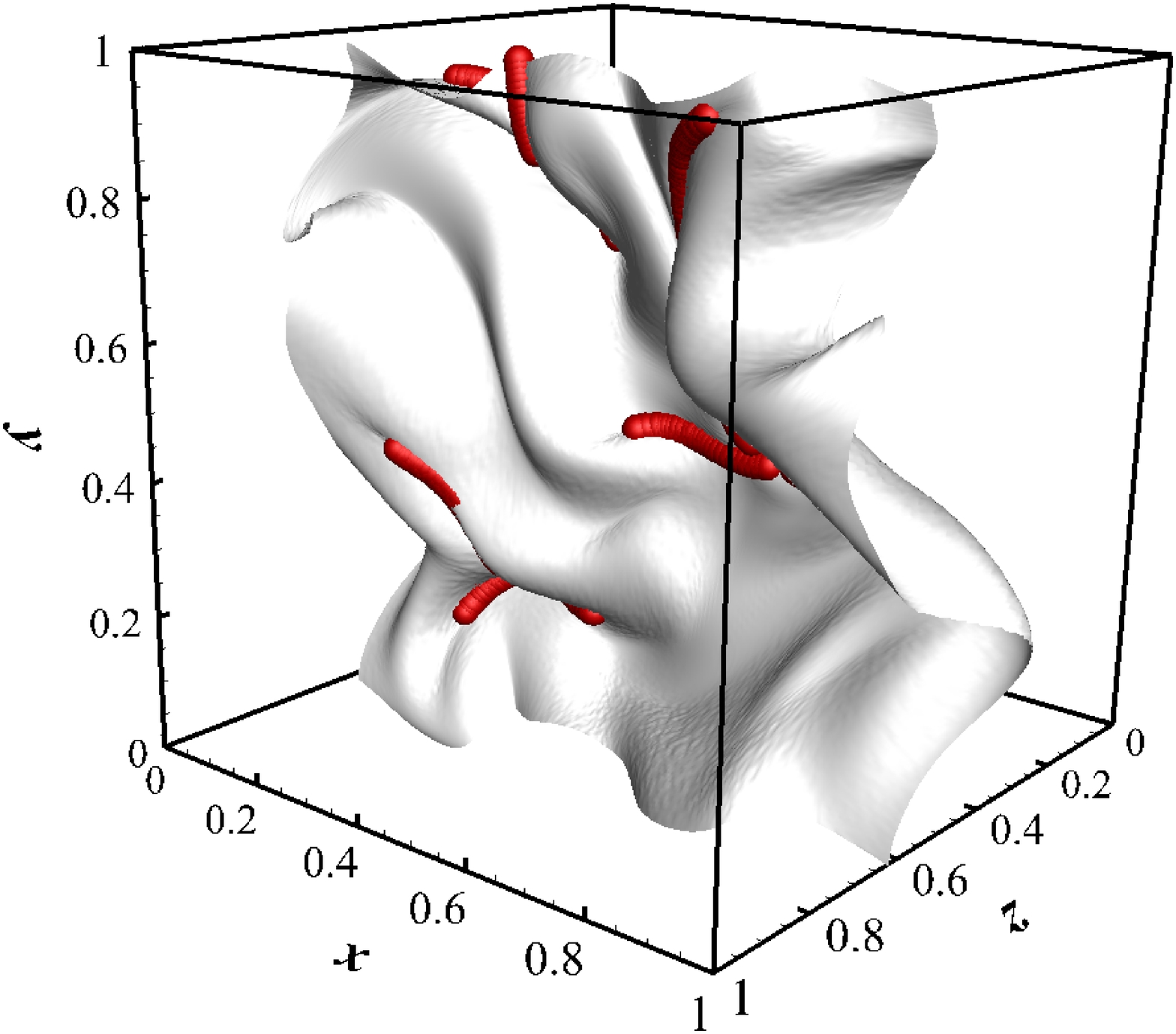}}
 \subfigure{
 \centering
 \includegraphics[width=0.4\textwidth]{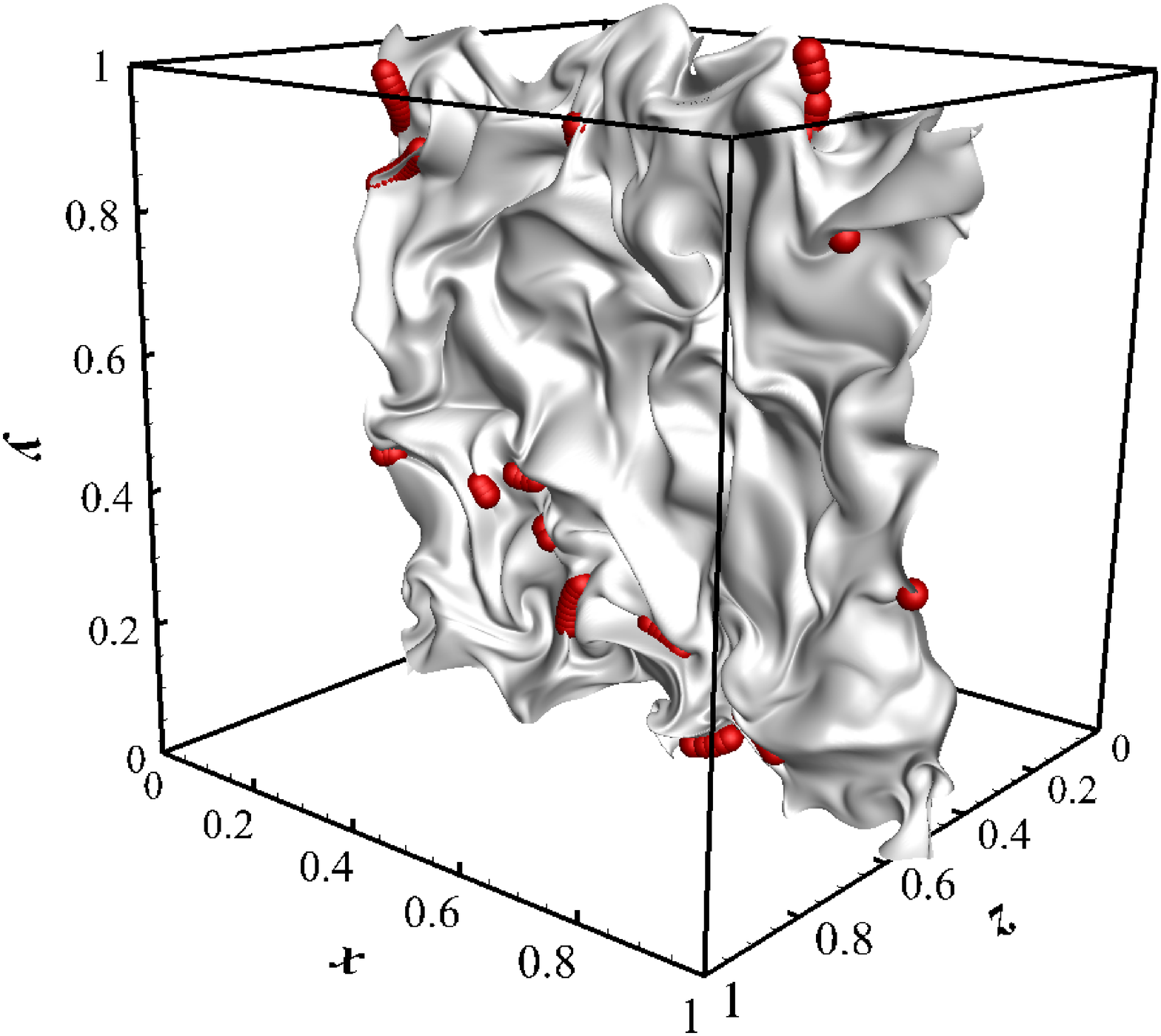}}

 \setcounter{subfigure}{0}
 \subfigure[Case B]{
 \centering
 \includegraphics[width=0.4\textwidth]{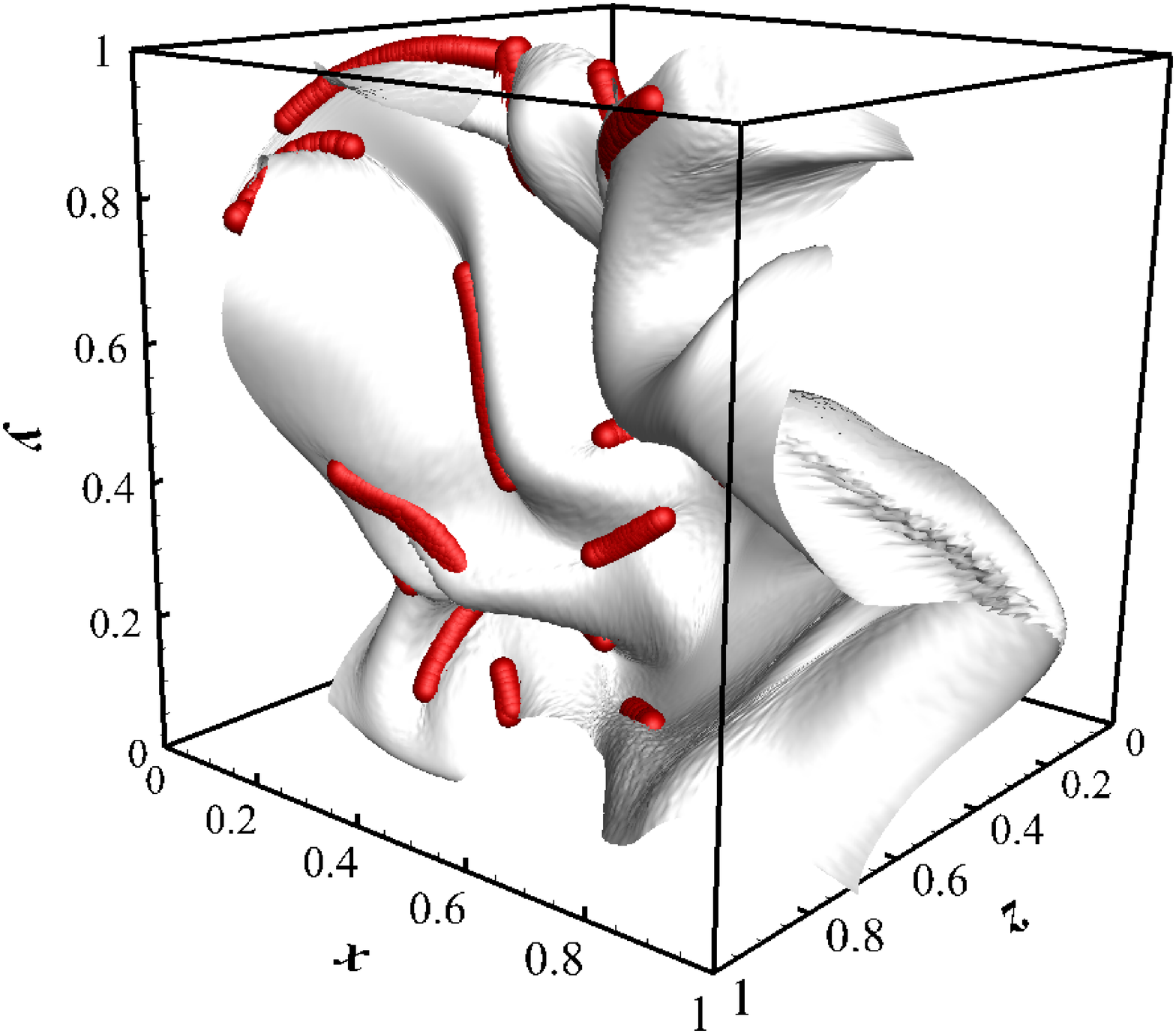}}
 \subfigure[Case D]{
 \centering
 \includegraphics[width=0.4\textwidth]{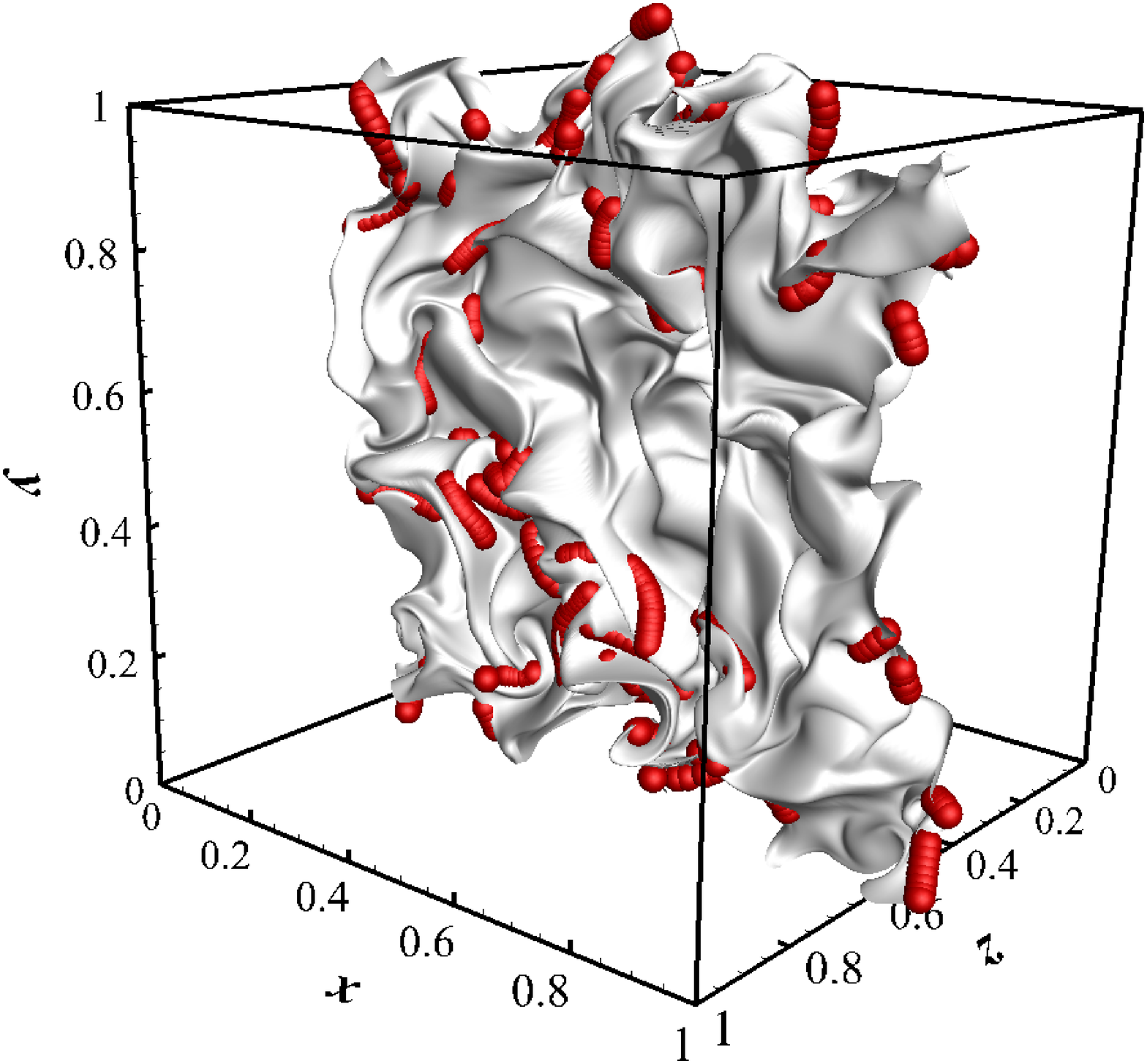}}
 \caption{(Colour online)~Temporal evolution of the global propagating surface in cases B and D at $t^* = 1$, 2 and 3 (from top to bottom) in non-reacting HIT. The arrow denotes the propagation direction, and the red dots mark locations of cusp generation.}
 \label{fig:Evol_H2}
\end{figure}

The evolution of global propagating surfaces consisting of surface elements in stationary HIT is shown in figure~\ref{fig:Evol_H2} for cases B and D. From an initial plane, the propagating surface is gradually wrinkled and corrugated by turbulent straining motion to form the cellular geometry. Compared with the persistent stretch of a passive material surface with $S_d=0$, the evolution of a propagating surface with finite $S_d$ can generate cusps, and cusp locations are marked by red dots.
Most of the cusps are generated around the crest of the surface with large negative curvatures. This `pocket structure' is critical for predicting the flame stabilization and pollution formation in turbulent premixed combustion \citep[][]{Bell2013}.
At the same normalized time $t^*\equiv t/\tau_\eta$, the surface is more wrinkled in case D with larger turbulence intensity than in case B.

We remark that the ensemble of surface elements can only approximate the global propagating surface rather than exactly represent the real flame surface. The surface elements can be considered as sample points scattered on the global propagating surface.
\rtwo{Moreover, the heat release in combustion can induce fluid thermal expansion and acceleration \citep[][]{Peters2000,Sabelnikov2017}. The relevant density change and pressure effects can influence the turbulent flame morphology and dynamics, such as cusp formation and collision events \citep[][]{Fogla2015}, and they can be coupled with the Darrieus--Landau instability under weak turbulence \citep[][]{Creta2011,Troiani2015}.
%
These variable-density effects are not considered in the present non-reacting HIT, and they can be incorporated via a modelled variable-density flow in future work.}

\section{Model development}\label{sec:mdoeldevelop}

\subsection{Turbulent burning velocity}

The turbulent burning velocity is defined using the global consumption speed of fuel
\EQ
S_T\equiv\frac{1}{\rho_{\textrm{u}} Y_{\textrm{f,u}} L_y L_z} \int_{\Omega} - \rho \dot{\omega}_{\textrm{f}}(x,y,z,t) \rd\Omega,
\label{eq:ST_DNS}
\EN
where $\rho_{\textrm{u}}$ denotes the density of the unburnt mixture, $\dot{\omega}_{\textrm{f}}$ the net reaction rate of fuel and $\Omega$ the entire computational domain.
Although there is no consensus on the best definition of $S_T$ \citep[][]{Driscoll2008}, i.e.~$S_T$ can be defined by the local consumption speed or the local displacement speed, we use the global consumption speed \eqref{eq:ST_DNS} without the subjective selection of the isocontour threshold.

\citet[]{Damkohler1940} conjectured that turbulence can enhance $S_T$ primarily through increasing the flame surface area under a low turbulence intensity, so the ratio of the flame surface area $A_T$ in turbulence and its projection in the propagating direction $A_0$ is generally equal to the ratio $S_T/S_L$.
Thus the modelling of $A_T$ is essential for estimating $S_T$ under the assumption \citep[see][]{Driscoll2008}
\EQ
\frac{S_T}{S_L} = I_0 \frac{A_T}{A_0}\approx \frac{A_T}{A_0}, 
\label{eq:ST-AT}
\EN
where the stretch factor $I_0$ depends on the effect of differential diffusion and is assumed to be unity from the present combustion DNS results (not shown).
Some other DNS studies of turbulent premixed flames \rtwo{in HIT} with moderate and high $u'$ \citep[see][]{Bell2002,Hawkes2006,Wang2017,Nivarti2017} also observe that $S_T/S_L$ are proportional to $A_T/A_0$ with nearly unity $I_0$, which supports the validity of Damk\"{o}hler's hypothesis \rtwo{for the flame propagation in HIT.
It is noted that in some jet flames under very strong turbulence \citep[\eg][]{Wabel2017}, \citet{Nivarti2019} argued that small-scale turbulence can serve as an effective turbulent diffusivity to enhance $S_T$, and this additional factor should be incorporated into \eqref{eq:ST-AT}.}

\subsection{Area growth of propagating surfaces} \label{sec:areagrowth}
\begin{figure}
    \centering
    \includegraphics[width=0.9\textwidth]{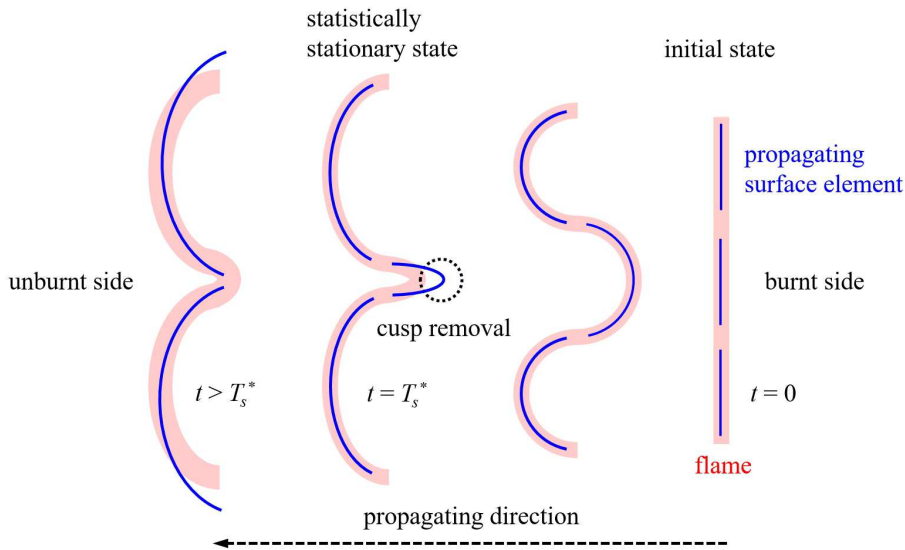}
    \caption{(Colour online)~The schematic diagram for the evolution of propagating surface elements (thin blue lines) and a flame front (thick red lines).} 
    \label{fig:sketch_Ts}
\end{figure}

We model the ratio $A_T/A_0$ of flame surface areas by the ratio of global propagating surface areas in the DNS of non-reacting HIT.
In the evolution of a global propagating surface initially consisting of $N_e$ surface elements, only $N_s(t)$ surface elements survive.
As sketched in figure~\ref{fig:sketch_Ts}, the other $N_d(t)$ surface elements disappear, because they evolve into cusps at a finite time and their surface areas shrink to zero \citep[][]{Girimaji1992}. Since these cusps cannot exist owing to smoothing effects of diffusion and curvature in real flames, these extreme samples are removed in the tracking of propagating elements \citep[see][]{Zheng2017}.
This removal of the cusps can be viewed as a mechanism of area reduction, and other flame destruction mechanisms, such as mutual annihilation and quenching, are implicitly involved in further modelling.

\begin{figure}
  \centering
  \psfrag{a}[c][c]{\footnotesize $t^*$}
  \psfrag{b}[c][c]{\footnotesize $R_s$}
  \psfrag{h}{\footnotesize Case A}
  \psfrag{i}{\footnotesize Case B}
  \psfrag{j}{\footnotesize Case C}
  \psfrag{k}{\footnotesize Case D}
  \psfrag{l}{\footnotesize Case E}
  \includegraphics[width=0.5\textwidth]{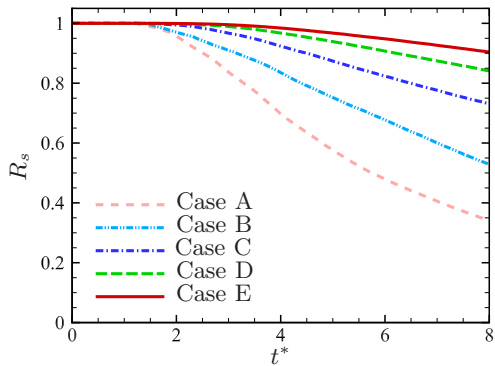}
  \caption{(Colour online)~Surviving ratios of propagating surface elements in non-reacting DNS cases.}
  \label{fig:Rs}
\end{figure}

The surviving ratio $R_s\equiv N_s(t)/N_e$ of surface elements is shown in figure~\ref{fig:Rs}.
We observe that the cusp generation occurs around $t^*=2$ and the cusps are generated more quickly at smaller turbulence intensity.
At smaller $u'/S_L$, the propagation term in \eqref{eq:curve} can dominate \citep[see][]{Zheng2017}, so the surface element with a large negative curvature is easier to develop a singularity in a finite time.

It is noted that all the subsequent quantities with the superscript `*' denote the ones normalized by Kolmogorov length scale $\eta$, time scale $\tau_\eta$ or velocity scale $u_\eta$.
\citet[]{Zheng2017} demonstrated that the profiles of ensemble-averaged $\kappa_1^*$ and $\kappa_2^*$ of propagating surfaces in terms of $S_d^*$ at the statistically stationary state are almost independent of $Re$.

After removing the cusps, the total surface area of the global propagating surface consisting of $N_s \le N_e$ surviving surface elements is approximated as
\EQ
A(t)\equiv \sum\limits_{N_s(t)}\delta A(t).
\EN
Summing up the local quantities of $N_s(t)$ surviving surface elements in (\ref{eq:area}) yields the evolution equation of $A(t)$ \citep[][]{Zheng2017}
\EQ
\de{}{t}\ln A(t) = {\<K_t(t)\>_A}  + 2\<S_d(t)\kappa(t)\>_A,
\label{eq:totalarea}
\EN
where
\EQ
\<f(t)\>_A \equiv \frac{\sum_{N_s(t)} f(t) \delta A(t)}{A(t)}
\label{eq:area_weighted_var}
\EN
denotes the area-weighted average of a quantity $f(t)$.

Integrating the normalized (\ref{eq:totalarea}) from the initial tracking time $t^*=0$ to a given time $t^*$ yields the surface area of the global propagating surface
\EQ
\frac{A(t^*)}{A_0} = \exp \left(\int_0^{t^*} \< K_t^*(s)\>_A \rd s\right)\exp\left(2\int_0^{t^*} \<S_d^*(s)\kappa^*(s)\>_A \rd s\right).
\label{eq:area2}
\EN
This implies that the growth of $A(t)$ of propagating surfaces is due to the Lagrangian history of statistically positive tangential strain-rate and propagation-curvature terms in the wrinkling process of propagating surfaces in turbulence \citep[][]{Zheng2017}. Subsequently $S_d$ is assumed to be a constant laminar burning velocity $S_L$ as
\EQ
\frac{A(t^*)}{A_0} = \exp \left(\int_0^{t^*} \< K_t^*(s)\>_A + 2S_L^*\<\kappa^*(s)\>_A \rd s\right).
\label{eq:area3}
\EN
This assumption is justified and the effect of flame stretch on $S_d$ is discussed in Appendix~\ref{sec:nonConstantSd}.

\begin{figure}
  \centering
  \psfrag{a}[c][c]{\footnotesize $t$~(ms)}
  \psfrag{b}[c][c]{\footnotesize $A(t)/A_0$}
  \psfrag{c}[c][c]{\footnotesize $t^*$}
  \psfrag{h}{\footnotesize Case A}
  \psfrag{i}{\footnotesize Case B}
  \psfrag{j}{\footnotesize Case C}
  \psfrag{k}{\footnotesize Case D}
  \psfrag{l}{\footnotesize Case E}
  \psfrag{m}{\footnotesize $\exp(\xi t^*)$}

  \subfigure{
  \centering
  \includegraphics[width=0.45\textwidth]{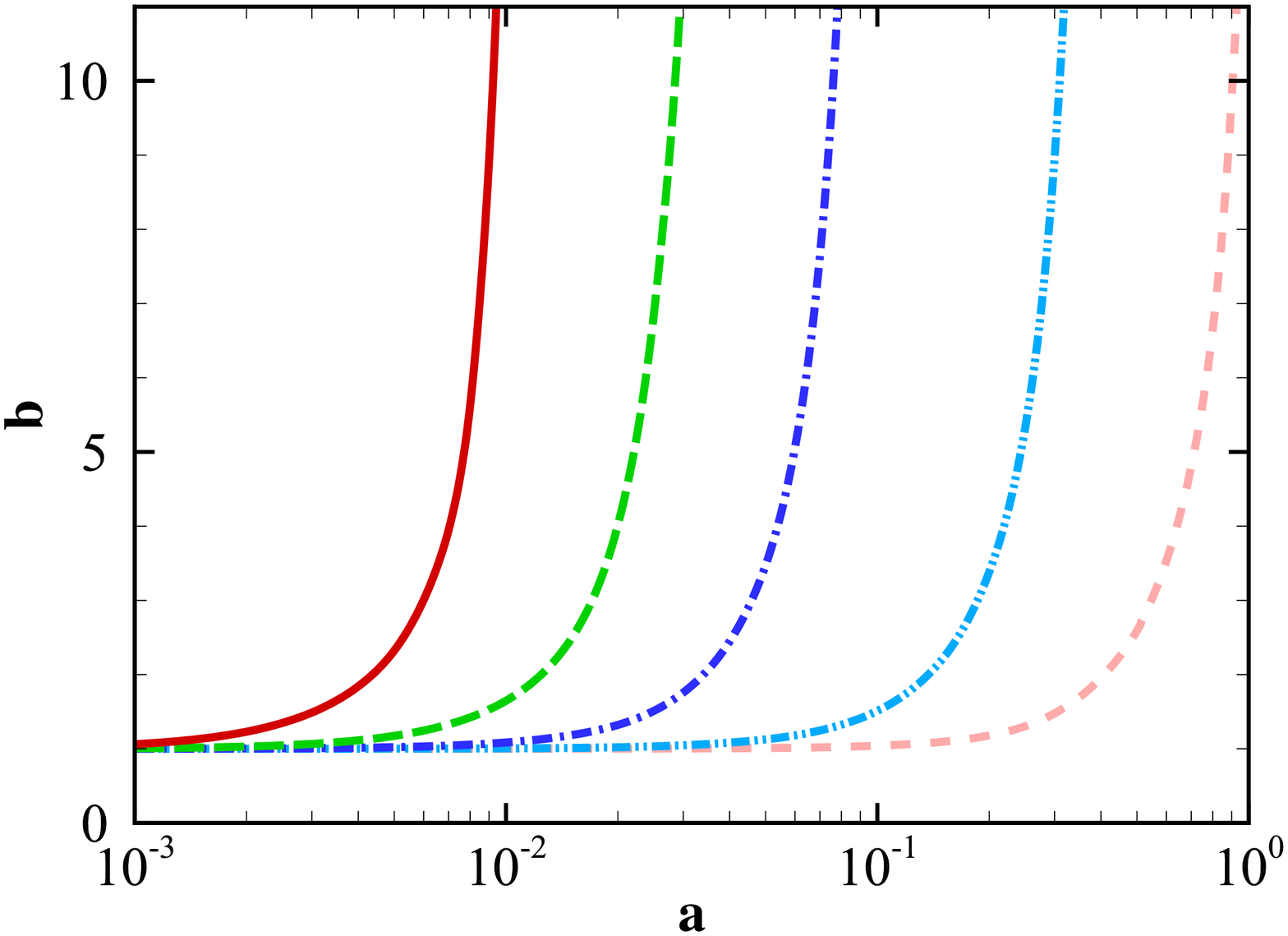}}
  \subfigure{
  \centering
  \includegraphics[width=0.45\textwidth]{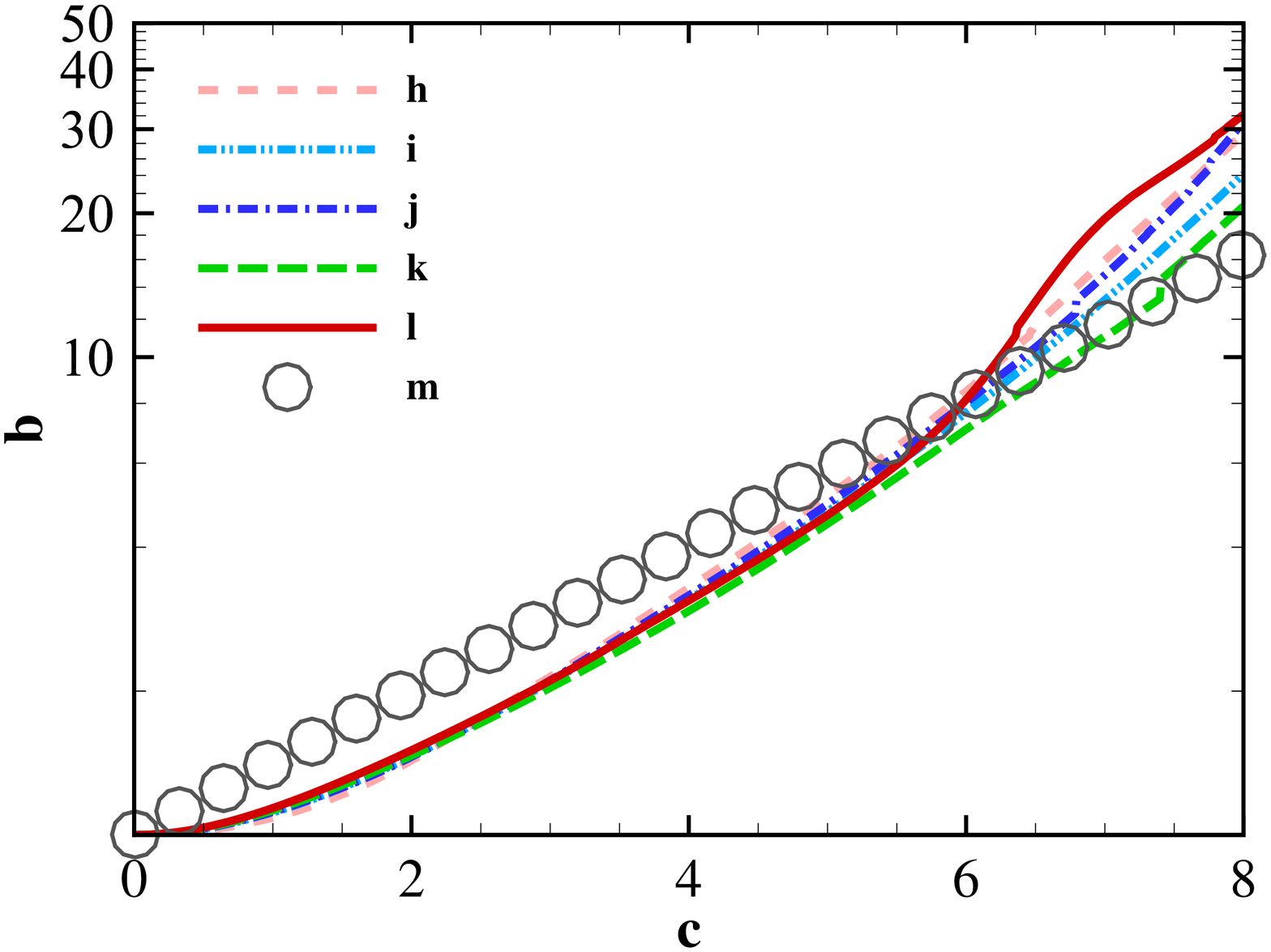}}

  \caption{(Colour online)~Surface area ratios of propagating surfaces in terms of (a) the physical time and (b) normalized time in non-reacting HIT.}
  \label{fig:AT_H2}
\end{figure}

As implied in \eqref{eq:area3}, $A(t)/A_0$ grows with $t$ in figure~\ref{fig:AT_H2}(a), and its growth rate increases with $Re$ for the same $S_L$, because $K_t$ scales as $K_t\sim 1/\tau_\eta \sim Re^{1/2}$.
This result is consistent with the observation that $A_T$ of flames increases with $Re$ in the argument of FSD modelling in Appendix~\ref{sec:FSD}.
On the other hand, if $t$ is normalized by corresponding $\tau_\eta$ in each case, the profiles of $A(t^*)/A_0$ for different $Re$ collapse in figure~\ref{fig:AT_H2}(b).
Similarly, the surviving ratio for the same $S_L^*$ is almost independent of $Re$ \citep[][]{Zheng2017}.
These indicate that Kolmogorov scales are more appropriate than integral scales for the normalization in scale analysis, and the further modeling based on the universal properties of $A(t^*)$ naturally involves the effect of $Re$ on $S_T$ in terms of integral quantities.

\begin{figure}
  \centering
  \psfrag{a}[c][c]{\footnotesize $t^*$}
  \psfrag{b}[c][c]{\footnotesize $\xi_A$}
  \psfrag{h}{\footnotesize Case A}
  \psfrag{i}{\footnotesize Case B}
  \psfrag{j}{\footnotesize Case C}
  \psfrag{k}{\footnotesize Case D}
  \psfrag{l}{\footnotesize Case E}

  \includegraphics[width=0.5\textwidth]{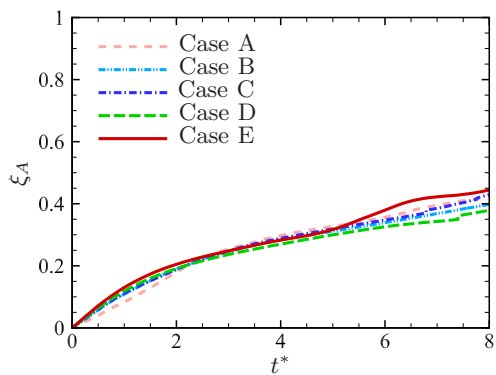}
  \caption{(Colour online)~Growth rates of the surface area of propagating surfaces in non-reacting HIT.}
  \label{fig:area_growth_rate}
\end{figure}

The estimation of $A(t)/A_0$ of propagating surfaces is similar to that for material surfaces in non-reacting HIT.
\citet{Batchelor1952} proposed that $A(t)/A_0$ of material surfaces grows exponentially owing to the persistent stretching of turbulent motion, which was validated by DNS \citep[\eg][]{Girimaji1990,Goto2007,Yang2010a}.
%
Considering the self-similar area growth of propagating surfaces at different $Re$, we re-express \eqref{eq:area3} into the form of
\EQ
\frac{A(t^*)}{A_0}=\exp(\xi_A t^*),
\label{eq:ATexp}
\EN
where $\xi_A(t^*)\equiv\log (A(t^*)/A_0 )/t^*$ is the area growth rate. The similar exponential growth of the flame surface was assumed by \citet{Kerstein1988} and \citet{Yakhot1988}.

In figure~\ref{fig:area_growth_rate}, the area growth rates also nearly collapse, consistent with the observation in figure~\ref{fig:AT_H2}(b).
For material surfaces in HIT, $\xi_A$ approaches a statistically stationary state as $\xi_A=0.33 \pm 0.04$ after a rapid, monotonic growth in $2 \sim 3 \tau_\eta$ \citep[see][]{Yang2010a}.
For propagating surfaces, by contrast, $\xi_A$ does not reach the statistically stationary state, so an additional approximation is introduced to model
\EQ
\xi_A(t^*)=\xi
\label{eq:xi_A}
\EN
by a constant $\xi$, which is addressed in \S\,\ref{sec:model_xi} in detail.

\subsection{Introduction of the truncation time}\label{sec:IntroTs}
The global propagating surface in turbulence should have $A(t)\rightarrow\infty$ as $t\rightarrow\infty$ in \eqref{eq:ATexp} owing to the persistent stretching and the lack of surface shortening mechanisms.
On the other hand, the shortening mechanism in real turbulent flames may occur when adjacent flamelets consume the intervening reactant, thereby annihilating both surface elements.
Thus the flame area $A_T$ should be finite, which can be considered as a stationary random variable with competing flame stretching and shortening mechanisms \citep[][]{Marble1977}.

To resolve this contradiction, we model the surface area of turbulent flames by
\EQ
A_T = A(T_s^*)
\label{eq:AT_Ts}
\EN
at a truncation time $T_s^*$ when the statistical geometry of propagating surfaces just reaches a statistically stationary state \citep[][]{Zheng2017}.
As sketched in figure~\ref{fig:sketch_Ts}, this state resembles the statistically equilibrium state in combustion between the flame area growth due to turbulent straining and the area consumption due to flame self-propagation. After this state, the nominal area of independent propagating surfaces with the infinitesimal thickness can still grow \rtwo(see figure~\ref{fig:AT_H2}), but the area of real flames statistically converges to a finite value.
%
%
Applying \eqref{eq:ATexp} and \eqref{eq:xi_A} to \eqref{eq:AT_Ts}, we have the model of $A_T$ in terms of $T_s^*$ as
\EQ
\frac{A_T}{A_0}= \exp \left( \xi T_s^* \right). 
\label{eq:ATmodel}
\EN

Before modelling $T_s^*$, we first use the characteristic curvature
\EQ
C^* \equiv \sqrt{ {\kappa_1^*}^2+{\kappa_2^*}^2 }
\EN
to characterize the statistical stationary state of propagating surfaces.
We find that \rtwo{the temporal evolution of the ensemble-averaged $C^*$ is statistically more robust than those of $K_t^*$ and $\kappa^*$ in \eqref{eq:area2}.}
The latter two are found to be very sensitive to specific realizations in DNS.

\begin{figure}
  \centering
  \psfrag{a}[c][c]{\footnotesize $t^*$}
  \psfrag{b}[c][c]{\footnotesize $\<C^*\>$}
  \psfrag{c}[c][c]{\footnotesize $d\<C^*\>/dt^*$}
  \psfrag{h}{\footnotesize Case A}
  \psfrag{i}{\footnotesize Case B}
  \psfrag{j}{\footnotesize Case C}
  \psfrag{k}{\footnotesize Case D}
  \psfrag{l}{\footnotesize Case E}

  \subfigure{
  \centering
  \includegraphics[width=0.45\textwidth]{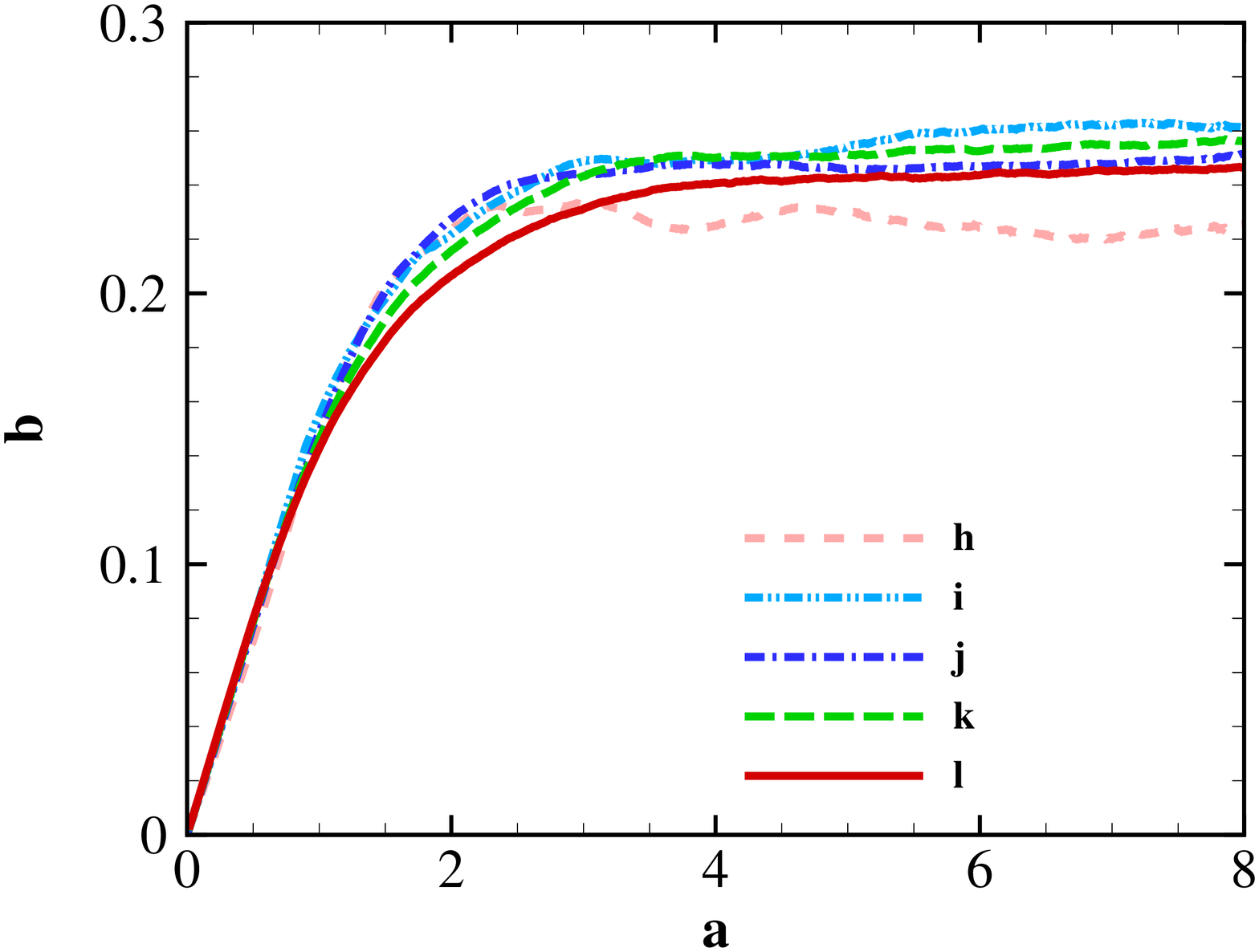}}
  \subfigure{
  \centering
  \includegraphics[width=0.45\textwidth]{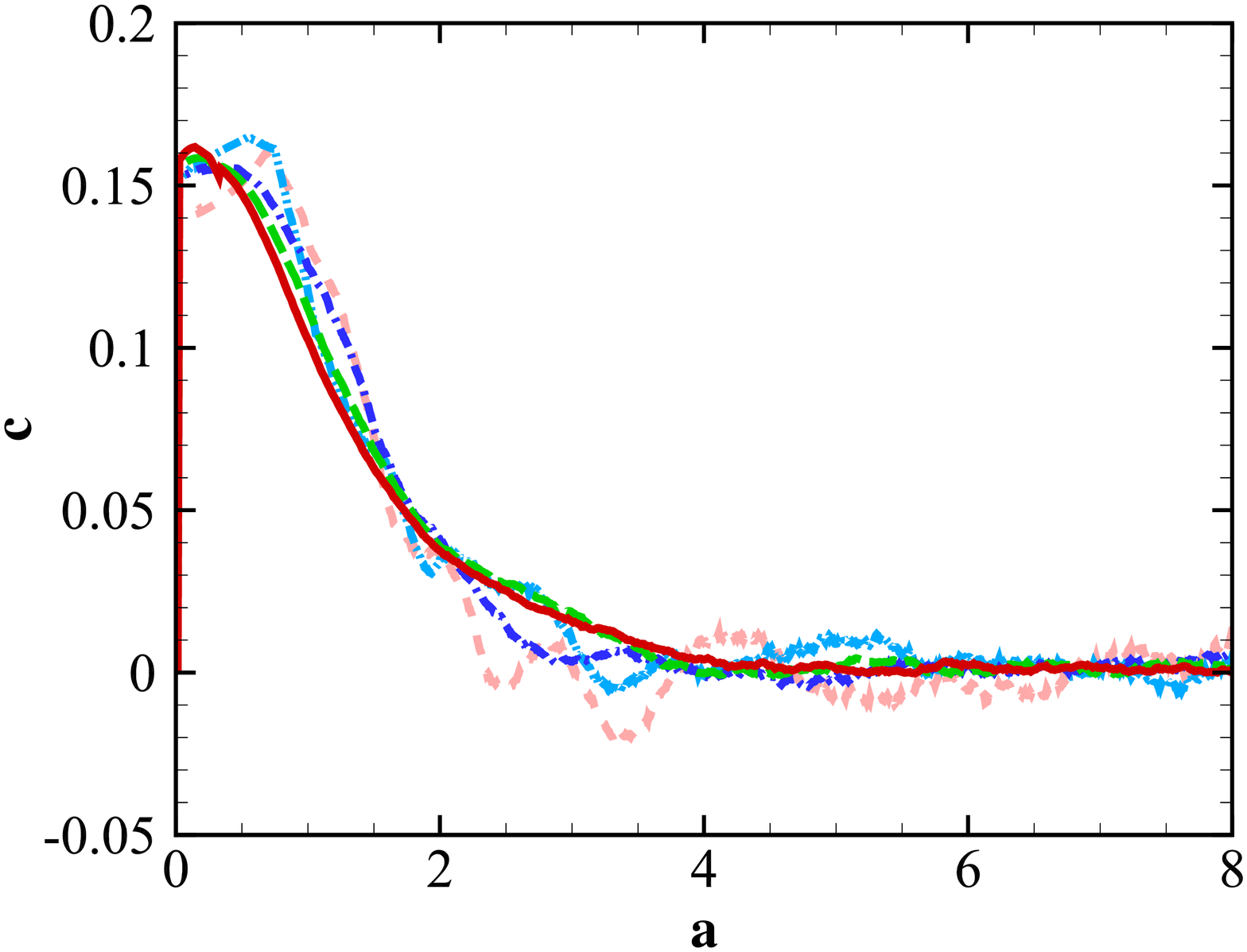}}
  \caption{(Colour online)~The temporal evolution of (a) the ensemble-averaged characteristic curvature and (b) its temporal growth rate of propagating surface elements in non-reacting HIT.}
  \label{fig:chf}
\end{figure}

\begin{figure}
  \centering
  \psfrag{a}[c][c]{\footnotesize $C^*$}
  \psfrag{b}[c][c]{\footnotesize PDF}
  \psfrag{c}{\footnotesize $t^*=2$}
  \psfrag{d}{\footnotesize $t^*=4$}
  \psfrag{e}{\footnotesize $t^*=6$}
  \psfrag{f}{\footnotesize $t^*=8$}
  \subfigure{
  \centering
  \includegraphics[width=0.45\textwidth]{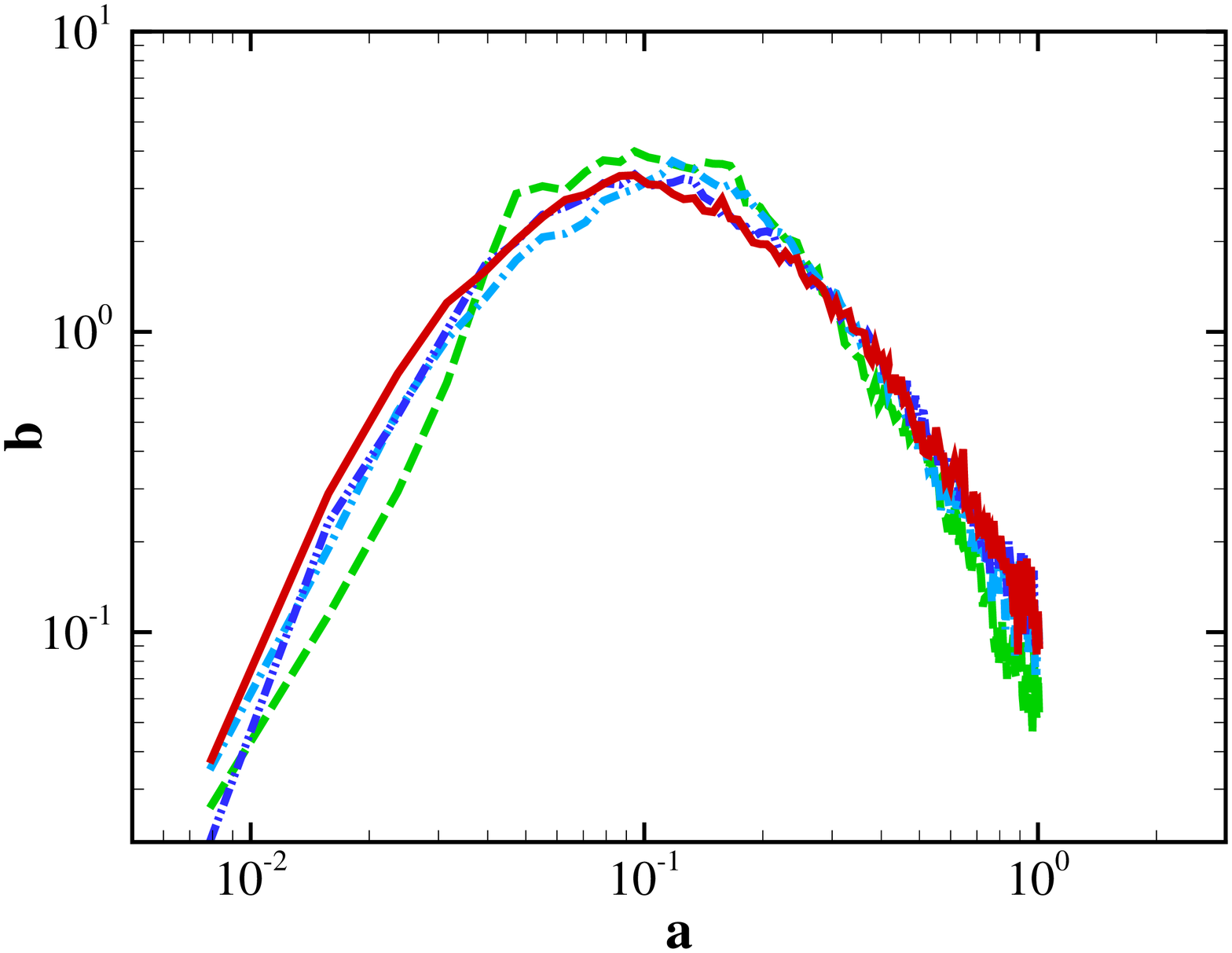}}
  \subfigure{
  \centering
  \includegraphics[width=0.45\textwidth]{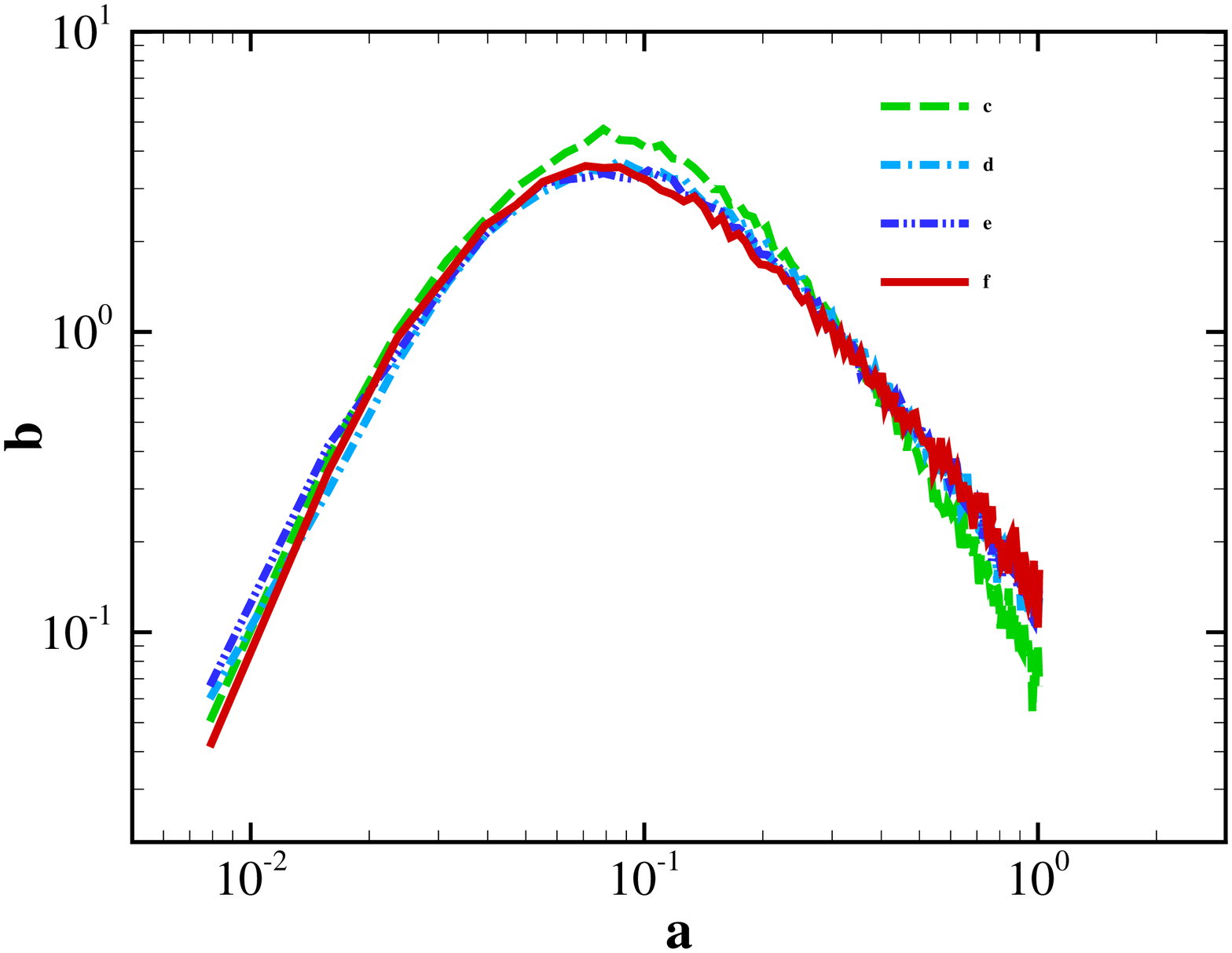}}
  \caption{\rtwo{(Colour online)~The temporal evolution of PDFs of the normalized characteristic curvature of propagating surfaces in non-reacting DNS (a) case B and (b) case D.}}
  \label{fig:chflogpdf}
\end{figure}

The DNS of propagating surfaces demonstrates that $\<C^*\>$ can reach a statistically stationary state after a short time in figure~\ref{fig:chf}(a), where $\<\cdot\>$ denotes the ensemble average over all the surviving surface elements, and this result is supported by the theoretical analysis in Appendix~\ref{sec:C_analysis}. The temporal growth rate $\rd\<C^*\>/\rd t^*$ of $\<C^*\>$ is also computed and shown in figure~\ref{fig:chf}(b). After a rapid growth at very small times, the growth rate decays and finally approaches a statistical stationary state. We observe that the growth rates almost collapse in cases C, D and E at moderate $Re$, which implies that the time $T_s^*$ for reaching the statistically stationary state of $\<C^*\>$ approaches to a finite value as $u'/S_L\rightarrow \infty$.
\rtwo{Furthermore, the temporal evolution of PDFs of $C^*$ in figure~\ref{fig:chflogpdf} indicates that the PDFs at different times almost collapse after reaching the statistically stationary state around $t^* > 4$.}

%

We denote the truncation time
\EQ
T_{\infty}^* = \lim_{u'/S_L\rightarrow \infty} T_s^*
\label{eq:Ts_inf}
\EN
for $u'/S_L\rightarrow \infty$ or material surfaces, which is useful in the modelling of $T_s^*$ in \S\,\ref{sec:Tsmodel}.
We approximate
\EQ
 T_{\infty}^*= \underset{t^*} {\arg \min} \left\{ \left| \de{\< C^*\>(t^*)}{t^*} \right| < \varepsilon_T\right\},
 \label{eq:Tinfty}
\EN
by $T_s^*$ from the DNS of propagating surfaces in case E with large $u'/S_L$, where $\varepsilon_T=0.1\%$ is a small threshold. 
We determine $T_{\infty}^* = 5.5$ using \eqref{eq:Tinfty} and it appears to be a universal constant, because the normalized time $t^*$ when $\<C^*\>$ of material surfaces reaches the stationary state \citep[][]{Girimaji1990,Girimaji1991} can be almost independent of large $Re$ owing to the self-similar statistical geometry.

\subsection{Modelling of the truncation time}\label{sec:Tsmodel}
The determination of the truncation time scale $T_s^*$ in \eqref{eq:ATmodel} is crucial for the modelling of $A_T/A_0$ using Lagrangian statistics of propagating surfaces, which is similar to the critical role of characteristic time scales in statistical turbulence models \citep[see][]{He2017}.
Based on discussions in \S\,\ref{sec:IntroTs}, $T_s^*$ is determined when averaged geometric quantities of propagating surfaces reach the statistically stationary state.
This state is an analogy to the statistically stationary $A_T/A_0$ of turbulent premixed flames in combustion DNS.

We observe that $T_s^*$ varies with $u'/S_L$ from the temporal evolution of $\<C^*\>$ in figure~\ref{fig:chf}, so we assume
\EQ
T_s^*=F\left(\frac{1}{S_L^*}\right)=F\left(Re^{-\frac{1}{4}}\frac{u'}{S_L}\right). 
\label{eq:Ts_func}
\EN
Next it is essential to determine the form of function $F$.
Inspired by the modelling approach for the width of a thermal wake in turbulent dispersion \citep[][]{Taylor1921} and the pressure--rate-of-strain tensor in Reynolds stress models \citep[][]{Launder1975,Pope2000}, we first determine $F$ at some limiting values of $u'/S_L$ with finite $Re$.

As $u'/S_L\rightarrow \infty$, propagating surfaces can be considered as material surfaces. According to the discussion in \S\,\ref{sec:IntroTs}, there exists a universal truncation time $T_{\infty}^*$ \eqref{eq:Ts_inf} when $\<C^*\>$ reaches the statistically stationary state.
If $F$ is a smooth function, \eqref{eq:Ts_inf} implies that
\EQ
\lim_{u'/S_L\rightarrow\infty}\de{T_s^*}{(1/S_L^*)}=0.
\label{eq:Ts_large}
\EN

As $u'/S_L\rightarrow 0$, we consider a laminar premixed flame with $u'=0$ and finite $S_L$. Since the initial planar propagating surface cannot deform, the truncation time is
\EQ
\lim_{u'/S_L \rightarrow 0}T_s^*=0.
\label{eq:Ts_0}
\EN

In weak turbulence with very small $u'/S_L$, the dependence of $S_T$ on $u'$ can be described by a general power law \citep[\eg][]{Peters1999,Creta2011,Lipatnikov2012}
\EQ
\frac{S_T}{S_L}=1+{\mc C} \left(\frac{u'}{S_L}\right)^{\zeta}, 
\label{eq:linear}
\EN
where $\mc C$ is independent of $u'$ but may depend on the unburnt mixture composition \citep[][]{Lipatnikov2002}, and $\zeta$ is an empirical constant.

The first-order Taylor expansion of the modelled area ratio \eqref{eq:ATmodel} with \eqref{eq:ST-AT} for very small $T_s^*$ and $u'/S_L$ is
\EQ
\frac{S_T}{S_L} = 1 + \xi T_s^*.
\label{eq:Taylor_expansion_ST}
\EN
It should be consistent with \eqref{eq:linear} at a laminar state with $Re = 1$, so we have
\EQ
\lim_{u'/S_L\rightarrow 0}\de{^n T_s^*}{(1/S_L^*)^n} = \frac{\mc C}{\xi}\prod_{m=0}^{n-1}(\zeta-m)\left(\frac{u'}{S_L}\right)^{\zeta-n}
\label{eq:Ts_small}
\EN
with an integer $1\le n\le\zeta$.
Here, we specify $\zeta=1$ as the simplest linear model
\EQ
S_T=S_L+\mc C u'
\label{eq:ST_linear}
\EN
in weak turbulence, which is generally applicable to various fuels.
Then \eqref{eq:Ts_small} is simplified as
\EQ
\lim_{u'/S_L\rightarrow 0}\de{T_s^*}{(1/S_L^*)} = \frac{\mc C}{\xi}.
\label{eq:Ts_small_linear}
\EN

In order to determine $\mathcal{C}$, we only need a very few sample data points
\EQ
G=\{(x_i, y_i)=(u'/S_L, S_T/S_L)\,|~ i \in \{0,1,...,N_G\} \}
\EN
from existing DNS or experimental measurement of propagating flames in weak turbulence with $u'/S_L < \epsilon_C$, where the small threshold value is set to $\epsilon_C = 2$ and the data point of $(x_0,y_0)=(0,1)$ at laminar condition is specified. Then, the least-squared fit is utilized to estimate $\mathcal{C}$ from $G$ as
\EQ
\mathcal{C} = \frac { \Sigma x_i y_i - \Sigma x_i \Sigma y_i /N_G} { \Sigma x_i ^ { 2 } -  ( \Sigma x_i )^2/N_G }.
\label{eq:fit_C}
\EN
This simple linear fit appears to be more robust than the high-order fit with $\zeta > 1$ in \eqref{eq:linear} from very sparse sample data points with finite uncertainties.

\begin{figure}
  \centering
  \includegraphics[width=0.5\textwidth]{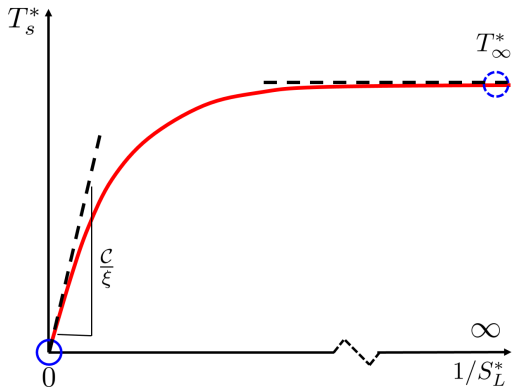}
  \caption{(Colour online)~A schematic diagram for the functional form of $F$ for $T_s^*$.
  Circles: values of $F$ at limiting conditions~\eqref{eq:Ts_inf} and \eqref{eq:Ts_0}; dashed lines: slopes of $F$ at asymptotic conditions~\eqref{eq:Ts_large} and \eqref{eq:Ts_small_linear}.}
  \label{fig:Ts_model}
\end{figure}

Considering the function $F$ at all the limiting values above (also sketched in figure~\ref{fig:Ts_model}), we propose a model for the truncation time as
\EQ
T_s^*=T_{\infty}^*\left[1-\exp\left(-\frac{\mc C}{\xi T_{\infty}^* {S_L^*}}\right)\right], 
\label{eq:Tsmodel}
\EN
which satisfies \eqref{eq:Ts_inf}, \eqref{eq:Ts_large}, \eqref{eq:Ts_0} and \eqref{eq:Ts_small_linear}, and is a monotonically increasing function in terms of $1/{S_L^*}$. \rtwo{It is noted that substituting the first-order Taylor expansion of \eqref{eq:Tsmodel} in terms of very small $1/S^*_L$ into \eqref{eq:Taylor_expansion_ST} with $Re=1$ recovers the linear model \eqref{eq:ST_linear}.}

\subsection{Modelling of the area growth rate}\label{sec:model_xi}
In the modelling of $A_T/A_0$ in \eqref{eq:ATmodel}, the constant $\xi$ needs to be determined by a function of given turbulence or flame parameters. Although $\xi_A$ grows with time in figure~\ref{fig:area_growth_rate}, we approximate $\xi_A$ as a constant effective growth rate
\EQ
\xi=\xi_A(t^*=T_{\infty}^*)
\label{eq:xi}
\EN
at the universal truncation time $T_{\infty}^*$.

Since the growth of $A(t^*)/A_0$ is independent of $Re$ as demonstrated in figure~\ref{fig:AT_H2}(b), $\xi$ at large $Re$ should be the same for a given mixture composition.
For example, we estimate $\xi=0.35$ by \eqref{eq:xi} from case E of DNS of propagating surfaces.
%
In figure~\ref{fig:AT_H2}(b), the comparison of $A(t^*)/A_0$ from DNS and the model \eqref{eq:ATexp} with $\xi=0.35$ indicates that \eqref{eq:ATexp} with \eqref{eq:xi} can provide a good approximation for the growth of $A(t^*)/A_0$.

Furthermore, $\xi$ may vary with $S_L^*$ for the same $Re$, so it is necessary to model this possible dependence of $\xi$ on different mixture compositions or fuels.
By comparing \eqref{eq:ATmodel} and \eqref{eq:area2}, we obtain
\EQ
\xi T_s^*=\int_0^{T_s^*} \<K_t^*\>_A+2S_L^*\<\kappa^*\>_A \rd s.
\label{eq:xiapprox}
\EN
Applying the mean value theorem for integrals to \eqref{eq:xiapprox} yields
\EQ
\xi = \frac{1}{2}\<K_t^*(t')\>_A+ S_L^*\<\kappa^*(t')\>_A,~t' \in (0, T_s^*),
\label{eq:xi_mean_value}
\EN
which indicates an explicit dependence of $\xi$ on the laminar flame speed.

\begin{figure}
  \psfrag{a}[c][c]{\footnotesize $S_{L0}$}
  \psfrag{b}[c][c]{\footnotesize $\xi$}
  \centering
  \includegraphics[width=0.5\textwidth]{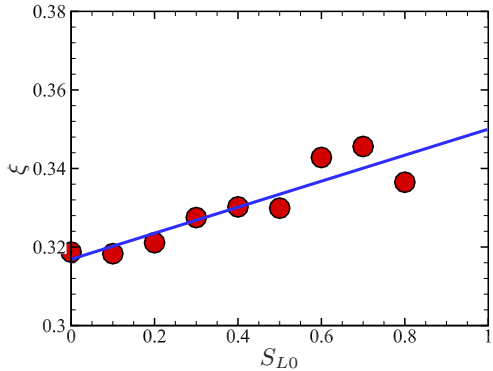}
  \caption{(Colour online)~Modelled area growth rates of propagating surfaces with $S_L=0\sim0.8$m/s in non-reacting HIT. Circles: results obtained by \eqref{eq:xi} from the tracking of propagating surfaces in case E; solid line: the linear fit \eqref{eq:ximodel} with $\mc A=0.317$ and $\mc B=0.033$.}
  \label{fig:ximodel}
\end{figure}

Considering the independence of $\xi$ for large $Re$, we use DNS case E of propagating surfaces with varying $S_L$ to fit \eqref{eq:xi_mean_value}.
We find that $\xi=0.33\pm0.02$ has a weak dependence on $S_L$ for a range of moderate $S_L$ in figure~\ref{fig:ximodel}, where the selection of $S_L=0\sim0.8$ m/s covers all the $S_L$ in the combustion DNS series discussed in \S\,\ref{sec:comparisonST}.
Since the dependence appears to be linear, we propose an empirical model
\EQ
\xi = \mc A + \mc B S_{L0}.
\label{eq:ximodel}
\EN
Here, model coefficients $\mc A=0.317$ and $\mc B=0.033$  are determined by the least-squared fit, and they represent the tangential strain-rate and propagation-curvature effects in \eqref{eq:xi_mean_value}, respectively; $S_{L0}=S_L/S_{L,\textrm{ref}}$ is a dimensionless laminar flame speed and is normalized by a reference value $S_{L,\textrm{ref}}=1$ m/s.

We remark that the model \eqref{eq:ximodel} can be valid for various fuels with moderate $S_L$, but it may break down for very large $S_L$.
In principle, $\xi$ of initial planar propagating surfaces is vanishing as $S_L/u'\rightarrow\infty$, so we speculate that $\xi$ may decrease at very large $S_L$.

\subsection{Predictive model of the turbulent burning velocity}\label{sec:modelremark}
Substituting models (\ref{eq:Tsmodel}) for $T_s^*$ and (\ref{eq:ximodel}) for $\xi$ into \eqref{eq:ATmodel} and using \eqref{eq:ST-AT}, we obtain
\EQ
\frac{S_T}{S_L}=\exp \left\{  T_{\infty}^* \left(\mc A + \mc B S_{L0}\right) \left[ 1- \exp \left( -\frac{\mc C Re^{-1/4}}{\left(\mc A + \mc B S_{L0}\right) T_{\infty}^*} \frac{u'}{S_L} \right)  \right]  \right\}.
\label{eq:STmodelf}
\EN
The physical meaning and determination method of the model parameters in \eqref{eq:STmodelf} are summarized in table \ref{tab:modelparameters}.
They are either given turbulence/flame parameters or universal constants pre-determined by Lagrangian statistics of propagating surfaces, except for $\mc C$ determined by \eqref{eq:fit_C} from one or a few available data points of $S_T$ of premixed flames in weak turbulence.

\begin{table}
	\centering
	\begin{tabular}{lccccccccccccccc}
		&Parameter   &Physical meaning    &Determination method     \\
		&$Re$        &Reynolds number of turbulence      & Given parameter          \\
		&$S_{L0}$    &Dimensionless laminar flame speed      & Given parameter          \\
		&$T_{\infty}^*$  &Time for the stationary state of m.s.     &Constant calculated from DNS of m.s.\\
		&$\mc A$      &Tangential strain-rate effect of p.s.      &Constant fitted from DNS of p.s.          \\
		&$\mc B$      &Propagation-curvature effect of p.s.      &Constant fitted from DNS of p.s.          \\
		&$\mc C$      &Combustion chemistry of fuel   &Fit from $S_T$ in weak turbulence \\
	\end{tabular}
    \caption{Summary of the model parameters in the proposed model \eqref{eq:STmodelf}, where m.s.~and p.s.~denote material and propagating surfaces in non-reacting HIT, respectively.}
	\label{tab:modelparameters}
\end{table}


The model \eqref{eq:STmodelf} can predict most of the basic trends of $S_T$ summarized in \citet{Lipatnikov2002}.
As $u'/S_L \rightarrow 0$ with $Re=1$, \eqref{eq:STmodelf} is reduced to the linear model \eqref{eq:ST_linear}, so the modelled $S_T$ increases with $u'$ and $\mc C$ for small $u'/S_L$.
The form of \eqref{eq:STmodelf} implies the `bending' of $S_T/S_L$ at moderate $u'/S_L$.
As $u'/S_L \rightarrow\infty$ with finite $Re$, this model converges to a finite value
\EQ
\frac{S_T}{S_L} = \exp \left[T_{\infty}^* \left(\mc A + \mc B S_{L0}\right)\right].
\EN
This implies that the modelled $S_T$ depends weakly on $u'$ and increases with the laminar flame speed at large $u'/S_L$, but it cannot predict the global quenching of flames characterized by a sharp drop of $S_T/S_L$ at very large $u'/S_L$ \cite[\eg][]{Karpov1978,Karpov1980,Driscoll2008,Lipatnikov2012}.

Substituting all the model constants $\mc A=0.317, \mc B=0.033$ and $T_\infty^*=5.5$ calculated from Lagrangian statistics of propagating or material surfaces in non-reacting HIT, we finally obtain a predictive model for the turbulent burning velocity in terms of $u'/S_L$ as
\EQ
\frac{S_T}{S_L}=\exp \left\{  \left( 1.742 + 0.182 S_{L0} \right) \left[ 1- \exp \left( -\frac{\mc C Re^{-1/4}}{1.742 + 0.182S_{L0}} \frac{u'}{S_L} \right)  \right]  \right\}.
\label{eq:STmodel_predict}
\EN
This model is validated in \S\,\ref{sec:comparisonST} by three DNS data sets of various fuels.

Compared with existing models of $S_T$, the features of the present one are summarized as follows. (1) The Lagrangian history of flame wrinkling is represented by the Lagrangian statistics of propagating surfaces in non-reacting HIT, and is utilized to construct the model. In other words, the model \eqref{eq:STmodelf} not only depends on local values of $u'$, flow integral property $Re$ and flame properties $S_L$ and $\mc C$, but also on the Lagrangian-based parameters $T_{\infty}^*$, $\mc A$ and $\mc B$.
(2) The $Re$-independent profile of $A(t^*)$ in terms of the quantities normalized by Kolmogorov scales is used in the modelling procedure, which introduces the universal statistics of propagating surfaces in small-scale turbulence and naturally involves the effect of $Re$ on $S_T$ in terms of integral scales.
(3) All the model parameters are derived via scale analysis and asymptotic analysis, and their values are determined by DNS of propagating surfaces in non-reacting HIT and existing data of $S_T$ in weak turbulence. 


\section{Model assessment} \label{sec:assessST}

\subsection{\textit{A posteriori} test} \label{sec:comparisonST}

We assess the model \eqref{eq:STmodel_predict} of $S_T$ by comparing the model prediction and the results from the present combustion DNS and another two DNS series in the literature \citep[see][]{Nivarti2017,Lee2012} with the same flame configuration in figure~\ref{fig:config} but with different fuels and $S_L$.

The three DNS series of turbulent premixed flames are briefly reviewed below.
(1) The present DNS: H$_2$/air flame with detailed chemistry.
(2) DNS of \citet{Nivarti2017}: CH$_4$/air flame with the chemistry described using a single-step Arrhenius expression.
(3) DNS (Group T) of \citet{Lee2010}: The premixed mixture is modelled by a reaction progress variable, and the chemistry is computed by the single-step Arrhenius expression.
%
As shown in figure~\ref{fig:regime}, these three DNS series cover a range of $Re$, $Ka$ and premixed combustion regimes.
All the model parameters in \eqref{eq:STmodel_predict} are summarized in table~\ref{tab:parampredict}, where $\mc C$ for each case is calculated from one or two corresponding DNS data points of $S_T$ at $u'/S_L<2$ by \eqref{eq:fit_C}.

\begin{table}
  \centering
  \setlength{\tabcolsep}{3.0mm}{
  \begin{tabular}{lcccc}
   &DNS series          &Present DNS            &\cite{Nivarti2017}         &\cite{Lee2010}   \\
   &Unburnt mixture     &H$_2$/air              &CH$_4$/air                 & Progress variable              \\
   &Chemistry           &Detailed               &One-step                   & One-step              \\
   &$S_{L0}$            &0.727                  &0.39                       & 0.0016              \\
   &$\mc C$             &2.43                   &0.55                       & 1.0             \\
  \end{tabular}}
  \caption{Summary of the model parameters in three DNS series with various fuels.}
  \label{tab:parampredict}
\end{table}

\begin{figure}
  \begin{center}
  \psfrag{x}{\footnotesize $u'/S_L$}
  \psfrag{y}{\footnotesize $S_T/S_L$}
  \psfrag{a}{\footnotesize{Present DNS}}
  \psfrag{b}{\footnotesize{\cite{Nivarti2017}}}
  \psfrag{c}{\footnotesize{\cite{Lee2010}}}
  \includegraphics[width=0.7\textwidth]{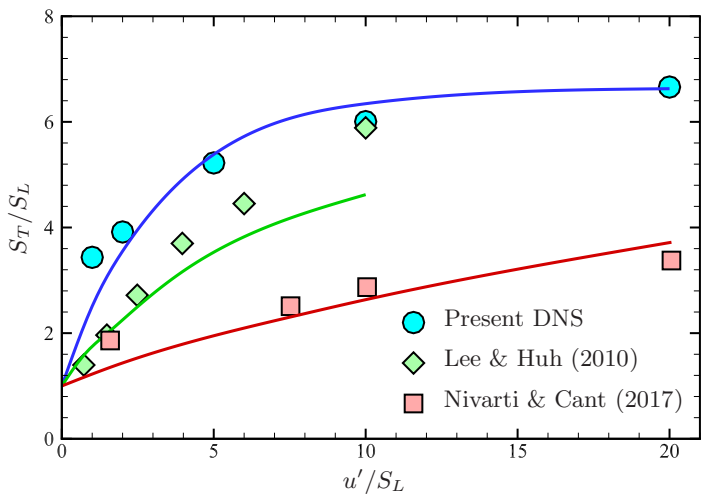}
  \caption{(Colour online)~Comparison of $S_T$ calculated from the combustion DNS (symbols) and the proposed model \eqref{eq:STmodel_predict} (solid lines).
  }
  \label{fig:STmodel}
  \end{center}
\end{figure}

The model \eqref{eq:STmodel_predict} is validated by combustion DNS of various fuels in figure~\ref{fig:STmodel}.
In general, the model predictions (solid lines) from \eqref{eq:STmodel_predict} agree well with the DNS results (symbols) and capture the bending of the profile of $S_T/S_L$ in terms of $u'/S_L$.
The overall good agreement ranges over $0<u'/S_L \leq 20$ and various fuels, indicating the generality of our proposed model.
In appendix~\ref{sec:model_comp}, the present model is also compared with other models of $S_T$ in the three DNS series, and the present one gives the overall best performance.
The discrepancies at very large $u'/S_L$ for DNS series of \cite{Nivarti2017} and \cite{Lee2010} are perhaps due to the breakdown of flame fronts close to the broken reaction zone. \rone{Moreover, uncertainties of model constants in \eqref{eq:STmodelf} can result in the discrepancies, which are elaborated below.}
%

\rone{First, the value of $\mc C$ is fuel-dependent and relies on available data points of $S_T$ under weak turbulence. In principle, the linear scaling \eqref{eq:ST_linear} is recovered at $u'/S_L=0$ with $Re\sim O(1)$.
On the other hand, if only the data points of $S_T/S_L$ with finite $u'/S_L$ and $Re > O(1)$ available, $\mc C$ fitted from \eqref{eq:fit_C} can cause the discrepancy from \eqref{eq:ST_linear} near $u'/S_L=0$ in figure~\ref{fig:STmodel}.}
%

\begin{figure}
  \begin{center}
  \psfrag{a}[c][c]{\footnotesize $u'/S_L$}
  \psfrag{b}[c][c]{\footnotesize $S_T/S_L$}
  \psfrag{c}{\footnotesize {Present DNS}}
  \psfrag{d}{\footnotesize $\mc C=1.5$}
  \psfrag{e}{\footnotesize $\mc C=2.0$}
  \psfrag{f}{\footnotesize $\mc C=2.5$}
  \psfrag{h}{\footnotesize {\cite{Nivarti2017}}}
  \psfrag{i}{\footnotesize $\mc C=0.5$}
  \psfrag{j}{\footnotesize $\mc C=1.0$}
  \psfrag{k}{\footnotesize $\mc C=1.5$}
  \psfrag{l}{\footnotesize {\cite{Lee2010}}}
  \subfigure{
  \centering
  \includegraphics[width=0.45\textwidth]{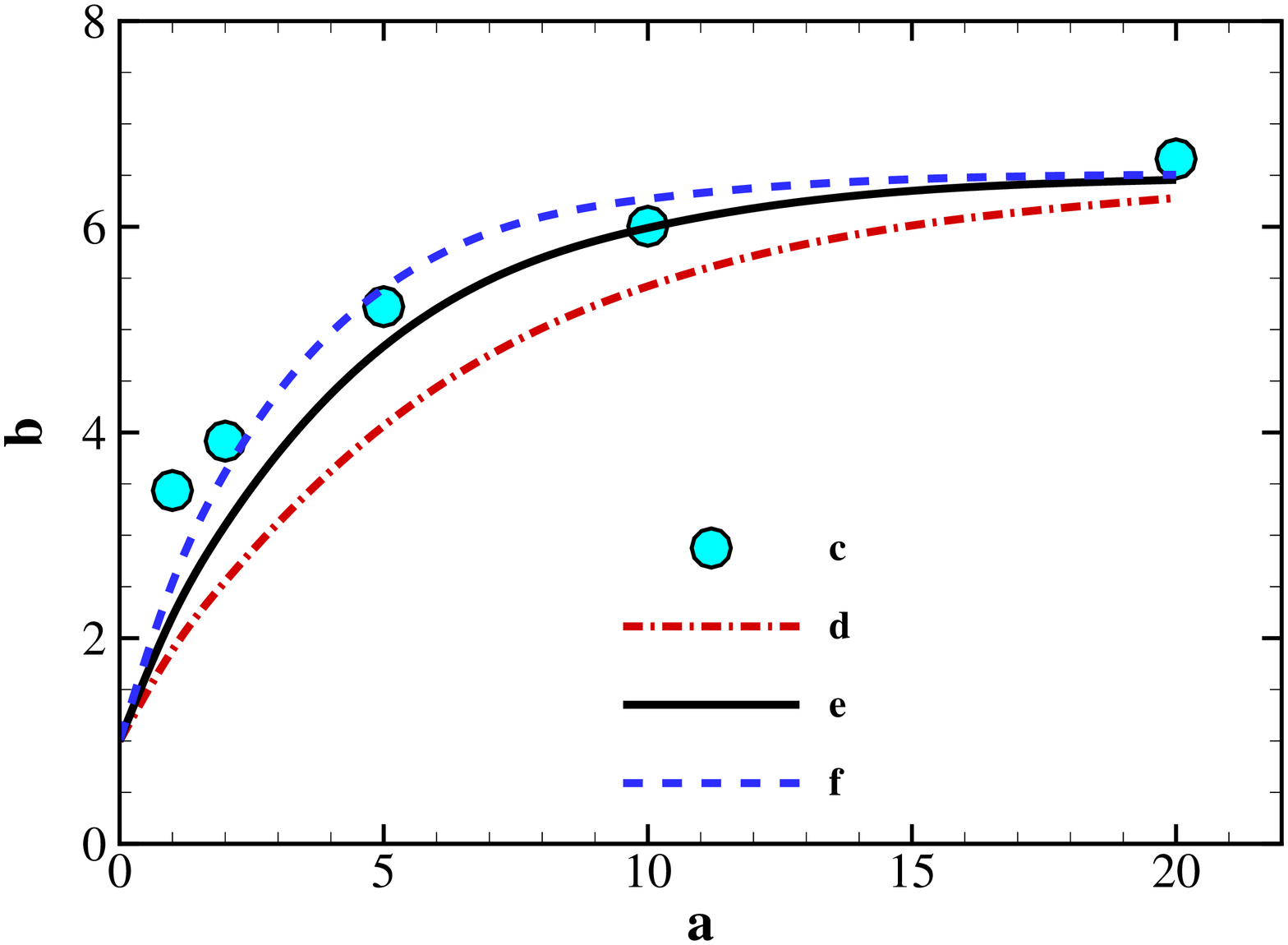}} \\

  \subfigure{
  \centering
  \includegraphics[width=0.45\textwidth]{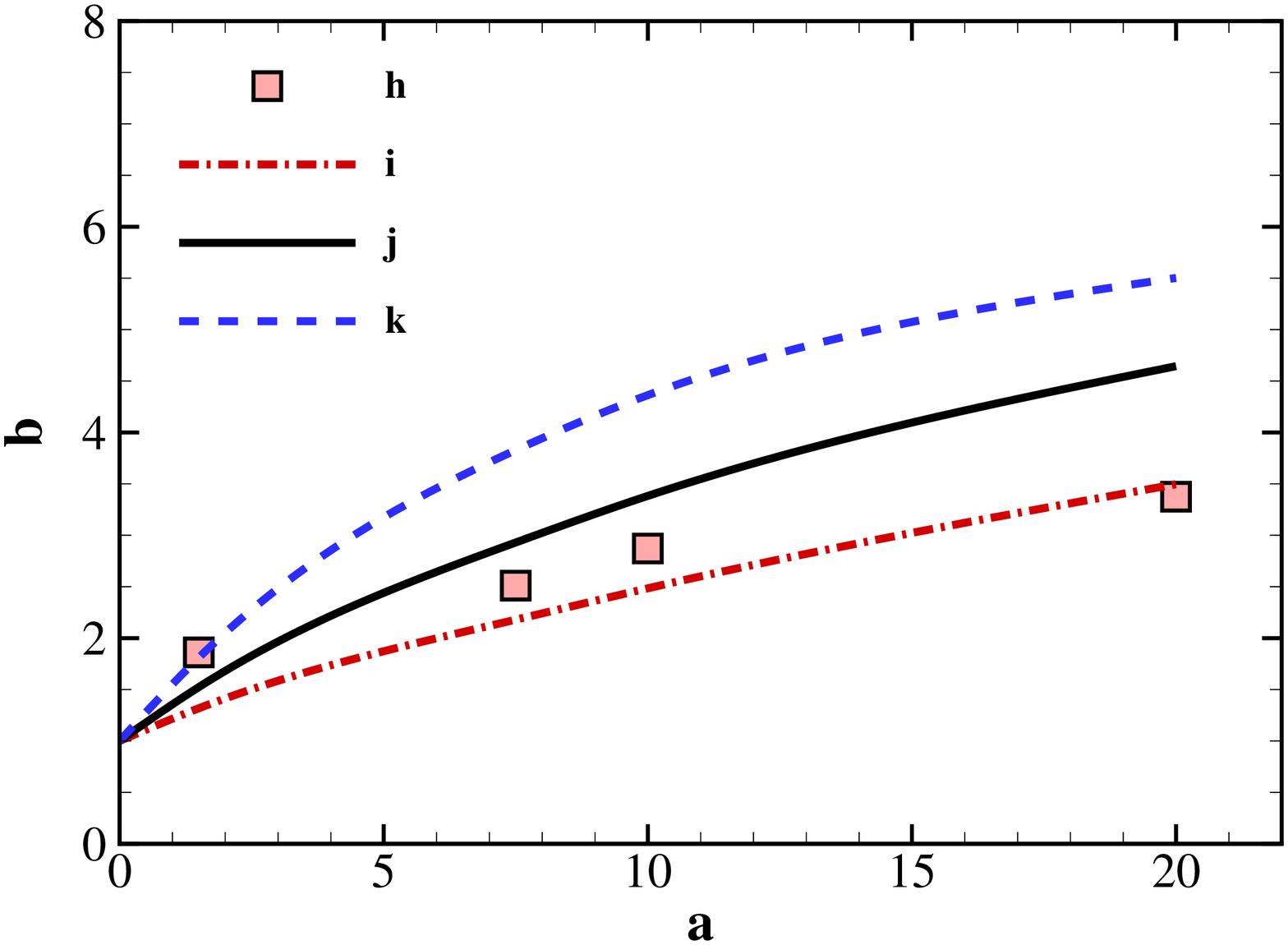}}
  \subfigure{
  \centering
  \includegraphics[width=0.45\textwidth]{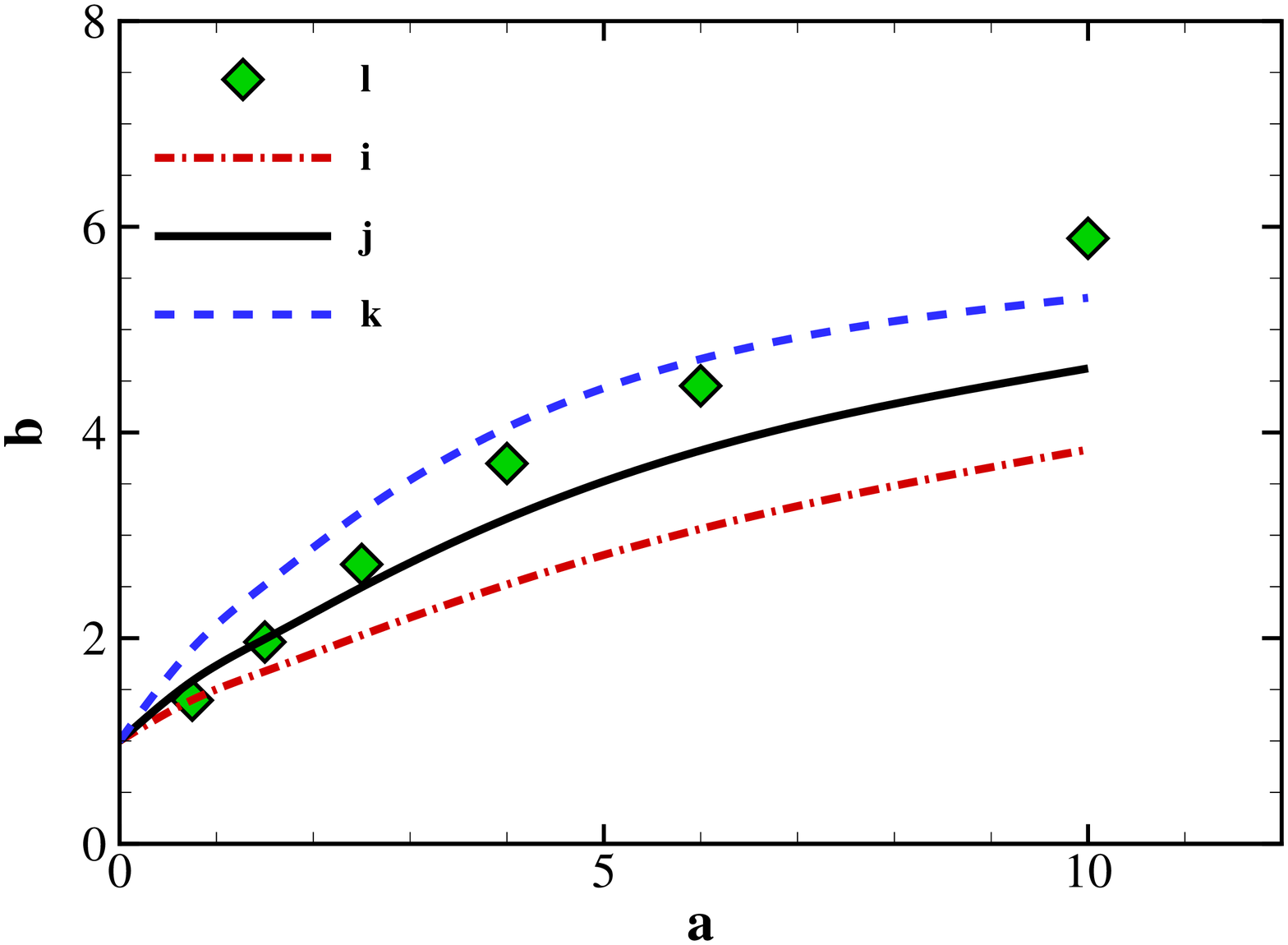}}
  \caption{\rtwo{(Colour online)~Model predictions of $S_T$ with the model constant $\mc C = \mc C_0 $ or $\mc C = \mc C_0\pm0.5$.}}
  \label{fig:Sensitivity_C}
  \end{center}
\end{figure}

\rtwo{Furthermore, if there is no available data of $S_T/S_L$ for fitting $\mc C$, we suggest that the value of $\mc C$ can be simply set to $\mc C_0=2.0$ for hydrogen fuels and $\mc C_0=1.0$ for other fuels based on the chemical activity of fuels. We observe the good agreement between the DNS result and the model prediction of \eqref{eq:STmodel_predict} with the empirical constant $\mc C=\mc C_0$ instead of fitting $\mc C$ from data, and illustrate the sensitivity of the model prediction on $\mc C$ by varying $\mc C = \mc C_0 \pm 0.5$ in figure~\ref{fig:Sensitivity_C}. In general, $S_T/S_L$ predicted from \eqref{eq:STmodel_predict} increases with $\mc C$. }


%
%

\rone{Second, the model constants $\mc A$, $\mc B$ and $T_\infty^*$ appear to be universal, but the determination of their values may involve some uncertainties owing to numerical schemes, forcing methods and finite-$Re$ effects in the DNS of non-reacting HIT and fitting methods for determining the constants, which can slightly affect the model prediction of $S_T$.}

\rtwo{Although the present $S_T$ model still has some uncertainties and discrepancies owing to the physical assumptions in model development and the lack of rigorous methods for determining $\mc C$ generally controlled by combustion chemistry, it is able to predict the relatively accurate bending effect for flames in HIT at large $u'/S_L$ in figures~\ref{fig:STmodel} and \ref{fig:Sensitivity_C}, which is superior to existing $S_T$ models (see detailed comparisons in Appendix~\ref{sec:model_comp}).}
In particular, the propagating surface used in the model was introduced only for the flamelet regime \citep[see][]{Girimaji1992}, but the present modelling approach appears to extend the concept of propagating surfaces to the thin-reaction-zone regime.
The truncation of the area growth by $T_s^*$ and the cusp removal of propagating surfaces result in a finite $A(T_s^*)$, which can model the statistically stationary $A_T$ under competing mechanisms between the flame area growth and consumption.

\subsection{\textit{A priori} test} \label{sec:pdfanalysis}

In order to further validate the key modelling assumption \eqref{eq:ATmodel}, we assess the contributions to the growth of $A_T$ in the right-hand side of \eqref{eq:area2} by comparing $\<K_t^*(T_s^*)\>_A$ and $\<\kappa^*(T_s^*)\>_A$ of propagating surfaces in non-reacting HIT with $\<K_t^*\>_A$ and $\<\kappa^*\>_A$ of the isosurface $c=\hat{c}$ at the statistically stationary state in the present combustion DNS.
The probability density functions (PDFs) of $(K_t^*)_A$ and $(\kappa^*)_A$ from propagating surfaces and flames in two typical cases B and D are plotted in figure~\ref{fig:athmcomp}.
Here, based on \eqref{eq:area_weighted_var} and the FSD described in Appendix~\ref{sec:FSD}, $(f)_A= f \delta A /(A/N_s)$ and $(f)_A=f\Sigma'/\Sigma$ denote area-weighted quantities for propagating surfaces and flames, respectively, and $\<f\>_A$ is equal to the integration of the product of $(f)_A$ and its PDF in the sample space. Moreover, $K_t$ and $\kappa$ in combustion DNS are calculated by \eqref{eq:Kt_FSD} and \eqref{eq:kappa_FSD}.

\begin{figure}
  \centering
  \psfrag{a}[c][c]{\footnotesize $(K_t^*)_A$}
  \psfrag{b}[c][c]{\footnotesize PDF}
  \psfrag{c}[c][c]{\footnotesize $(\kappa^*)_A$}
  \subfigure{
  \centering
  \includegraphics[width=0.45\textwidth]{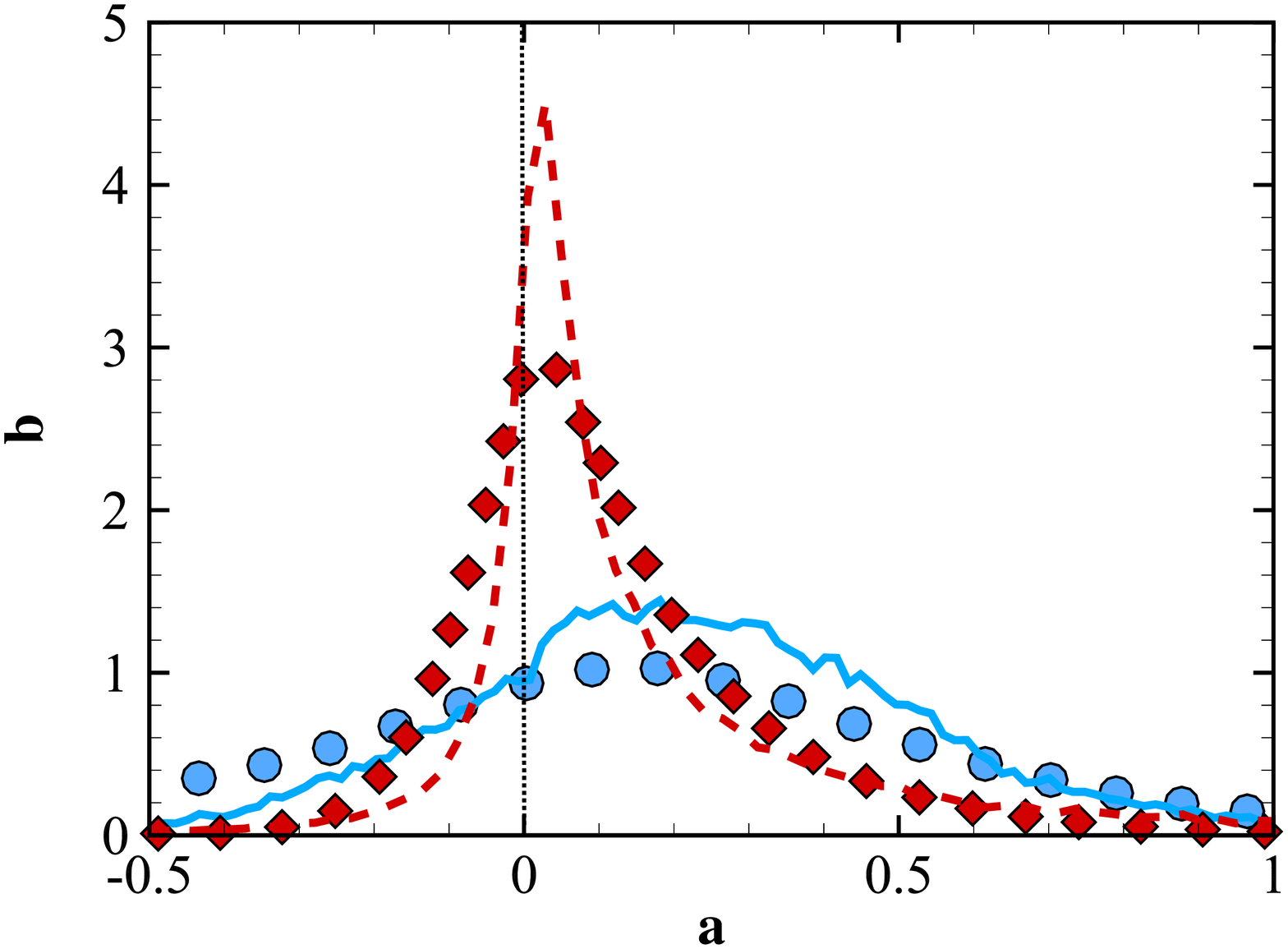}}
  \subfigure{
  \centering
  \includegraphics[width=0.45\textwidth]{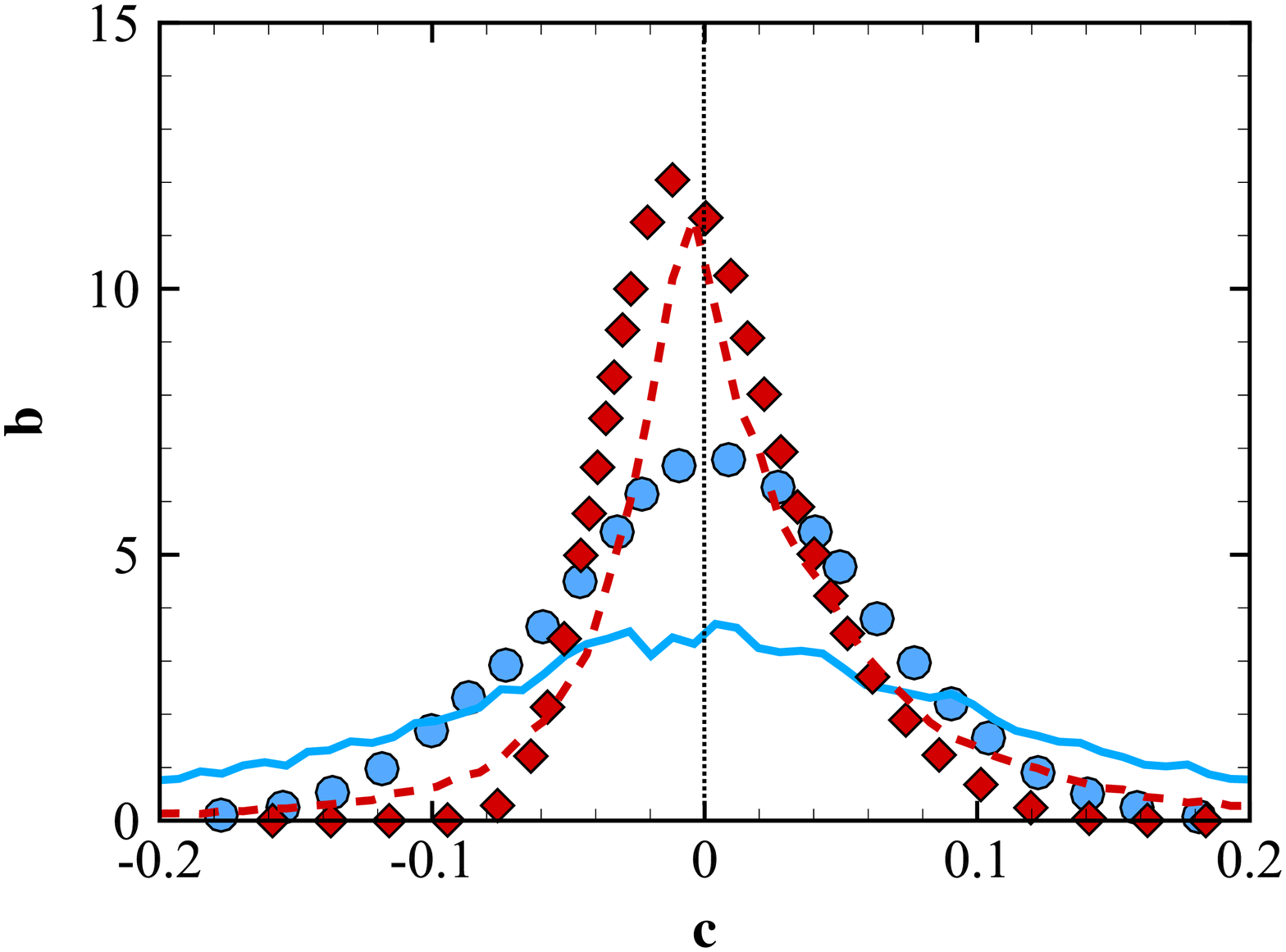}}
  \caption{(Colour online)~PDFs of (a) the normalized area-weighted tangential strain rate and (b) the mean curvature from flames and propagating surfaces in DNS cases B and D.
  Circles: flame in case B; diamonds: flame in case D; solid lines: propagating surface in case B; dashed lines: propagating surface in case D.}
  \label{fig:athmcomp}
\end{figure}

Figure~\ref{fig:athmcomp}(a) indicates that all the cases exhibit positive $\<K^*_t\>_A$, which is consistent with the positive contribution of turbulent straining to the area growth of material surfaces \citep[][]{Girimaji1990}.
The similar PDFs of $(K_t^*)_A$ in propagating surfaces and flames imply that the straining process experienced by propagating surfaces in non-reacting HIT resembles that of flames in turbulent combustion.

Figure~\ref{fig:athmcomp}(b) shows that the PDF of $(\kappa^*)_A$ of propagating surfaces is qualitatively similar to that of flames, but the former distribution is generally broader than the latter one. This observation implies that nearly singular propagating surfaces with very large curvatures can be smoothed in real flames owing to diffusion and curvature effects.
Furthermore, $\<\kappa^*\>_A$ of flames is close to zero, which agrees with the former finding \citep[\eg][]{Trouve1994,Chakraborty2005,Nivarti2017} that the curvature effect can consume the flame area.
On the other hand, $\<\kappa^*\>_A$ of propagating surfaces is slightly positive with the order of $O(0.01)$.
The cellular structures of propagating surfaces in figure~\ref{fig:Evol_H2} with positive $\<\kappa^*\>_A$ \citep[][]{Zheng2017} are only observed in turbulent premixed flames at low $Ka$  in figure~\ref{fig:FSD} \citep[see][]{Matalon2009}, whereas they break down at high $Ka$. This structural difference causes the discrepancy between PDFs of $(\kappa^*)_A$ for propagating surfaces and flames in case D.

\section{Conclusions}\label{sec:conclusions}

We propose a predictive model of $S_T$ in premixed combustion \rtwo{in HIT} based on the Lagrangian statistics of propagating surface elements in non-reacting stationary HIT. An ensemble of propagating surface elements initially constitutes a plane in the non-reacting HIT to mimic the propagation of a planar premixed flame front. The effects of molecular transport and reaction on each surface element is modelled via a constant displacement speed $S_d = S_L$.

The model development involves several assumptions and modelling steps.
First, $S_T/S_L$ is approximated by the ratio $A_T/A_0$ of flame surface areas from Damk\"{o}hler's assumption \eqref{eq:ST-AT}.
Then $A_T/A_0$ is approximated by the area ratio $A(t^*)/A_0$ of global propagating surfaces at the truncation time $t^*=T_s^*$ which signals that the characteristic curvature of initially planar propagating surfaces has reached the statistically stationary state in non-reacting HIT.
This state resembles the statistically equilibrium state in combustion between the flame area growth due to turbulent straining and the area consumption due to flame self-propagation.
In addition, we demonstrate that the temporal growth of $A(t^*)/A_0$ can be approximated by an exponential function \eqref{eq:ATmodel} with a constant growth rate which is modelled by a linear function \eqref{eq:ximodel} of the dimensionless $S_L$.
Accounting for $T_s^*$ at limiting conditions of very weak and strong turbulence, we propose the model \eqref{eq:Tsmodel} of $T_s^*$ by theoretical analysis and tracking of propagating surfaces.
Finally, we obtain the predictive model \eqref{eq:STmodel_predict} of $S_T$, an explicit expression in terms of $u'/S_L$ and several model parameters.
The model parameters include universal constants $T_\infty^*$, $\mc A$ and $\mc B$ calculated from Lagrangian statistics of propagating surfaces in non-reacting HIT, and a fuel-dependent $\mathcal{C}$ pre-determined by \eqref{eq:fit_C} from one or a few available combustion DNS or experimental data points of $S_T$ in weak turbulence.

Being different from other models for $S_T$, the present one incorporates the Lagrangian information of propagating surfaces characterizing the enhancement of the area growth of flames in turbulence and the universal statistics of propagating surfaces in small-scale turbulence involving the effect of $Re$.
Furthermore, the effect of fuel chemistry on $S_T$ is also considered by the model parameter $\mathcal{C}$.





The proposed model \eqref{eq:STmodel_predict} is validated by an \textit{a posteriori} test against three DNS series of turbulent premixed flames in HIT, including the present combustion DNS and the other two in the literature, which have the same flame configuration but different fuels and $S_L$. The model predictions of $S_T$ generally agree well with DNS results under a broad range of turbulent intensities and premixed combustion regimes.
The expression \eqref{eq:STmodel_predict} of the model implies that it can capture the basic trends of $S_T$ in terms of $u'$, i.e.~the linear growth in weak turbulence and the bending effect in strong turbulence.
Furthermore, the model validity is supported by an \emph{a priori} test.
We calculate PDFs of tangential strain-rate and propagation-curvature terms, which are two major contributions to the surface area growth, and find that the PDFs of each term in flames and propagating surfaces qualitatively agree.
%


On the other hand, the assumptions in the proposed model \rtwo{and the uncertainties in model constants} cause some limitations \rtwo{in the model formulation and discrepancies in the model prediction}, which can be improved in future work.
(1) Under very strong turbulence \citep[\eg][]{Wabel2017,Aspden2011}, the flame can be in the broken reaction zone in which no distinct flame surface exists, so the model assumption \eqref{eq:ST-AT} may be invalid.
(2) The model is developed based on the propagation of a planar flame \rtwo{in HIT}, so it may need to be adapted for different flame configurations or geometries \citep[][]{Driscoll2008}.
(3) The dependence of $S_T$ on pressure, gas expansion and differential diffusion \citep[see][]{Lipatnikov2012} is not considered in the present model.

\section*{Acknowledgments}
Y. Yang thanks Z. Ren, Z. Lu, N. Peng and S. Xiong for helpful comments. We gratefully acknowledge Caltech, the University of Colorado at Boulder and Stanford University for licensing the NGA code used in this work. This work has been supported in part by the National Natural Science Foundation of China (Grants No.~91541204, 91841302 and 11522215).

\appendix
\section{Modelling of the FSD}\label{sec:FSD}

In the modelling approach of FSD, the generalized local FSD $|\bs \nabla c|$ \citep[][]{Veynante2002} is employed to calculate the local surface-to-volume ratio $\Sigma' \equiv |\bs \nabla c|$. The generalized FSD $\Sigma\equiv\<\Sigma'\>_{c = \hat{c}}$ denotes the surface-averaged $\Sigma'$ along the isosurface of $c=\hat{c}$, i.e.~the flame surface-to-volume ratio in $\Omega$.

The evolution equation of $\Sigma$ \citep[see][]{Pope1988,Trouve1994} is
\EQ
\D{\Sigma}{t}+ \bs\nabla \bs\cdot\left(\< {\bs u} + S_d {\bs n}\>_A \Sigma\right) = \< K \>_A \Sigma,
\label{eq:sigma}
\EN
where $\bs n =-\bs \nabla c / |\bs \nabla c|$ denotes the flame normal, and $\< f \>_A \equiv \<f\Sigma'\>/\Sigma$ denotes the area-weighted average of a function $f$ over the isosurface of $c=\hat{c}$.
In the right-hand side of (\ref{eq:sigma}), the local flame stretch rate $K= K_t + 2S_d \kappa$ involves the tangential strain-rate term
\EQ
K_t=\bs\nabla {\bs u}-\bs{nn} \bs:\bs\nabla {\bs u}
\label{eq:Kt_FSD}
\EN
and the propagation-curvature term $2S_d \kappa$ with
\EQ
\kappa=\frac{\bs \nabla \bs\cdot \bs n}{2}.
\label{eq:kappa_FSD}
\EN
These expressions in Eulerian coordinates are consistent with the Lagrangian formulation in \eqref{eq:area}.
The turbulent flame area can be obtained by integrating the local FSD as
\EQ
A_T =\int_{\Omega} \Sigma'\delta (c-\hat{c}) \,\rd V.
\label{eq:ATmodel_FSD}
\EN

Since the turbulent burning velocity can be modelled by \eqref{eq:ST-AT} with \eqref{eq:ATmodel_FSD}, $S_T$ is generally increased with $u'/S_L$ as the increasing integration of the FSD throughout the flame brush \citep[][]{Nivarti2017}, but the growth of $S_T$ can slow down or even $S_T$ can decrease for large $u'/S_L$ if the clear flame front is destructed in very strong turbulence, as shown in case E in figure~\ref{fig:FSD}.

\section{Effects of non-constant $S_d$ on the area growth of propagating surfaces} \label{sec:nonConstantSd}
In the DNS of propagating surfaces in non-reacting HIT, the constant displacement speed $S_d=S_L$ can be modified by including the effects of flame stretch.
From the asymptotic theory, the first-order correction term for small curvature and strain rate is added as \citep[][]{Peters2000,Matalon2009}
\EQ
S_d=S_L-\mathcal{L} K= S_L(1-Ma Ka K^*),
\label{eq:nonSd}
\EN
where the Markstein number $Ma=\mathcal{L}/\delta_L$ denotes the ratio of the Markstein length $\mathcal{L}$ to the laminar flame thickness $\delta_L$ and measures the sensitivity of the flame speed to the local flame stretch.
For the H$_2$/air mixture with the equivalence ratio $\phi=0.6$ in the present combustion DNS, we obtain $Ma=-0.18$ \citep[see][]{Davis2002}.

To investigate the effects of non-constant $S_d$, \eqref{eq:nonSd} is applied to \eqref{eq:loc} in the tracking of propagating surfaces in non-reacting HIT.
The PDFs of $S_d$ of propagating surfaces are evaluated at the truncation time and are compared with the combustion DNS data.
Here, the displacement speed of the flame front, i.e.~the isosurface of $c=\hat{c}$, is calculated by \citep[][]{Pope1988}
\EQ
S_d=\frac{\dot{\omega}_c+ \bs \nabla \bs\cdot \left(\rho D_c \bs \nabla c\right)}{\rho | \bs \nabla c | },
\EN
where $\dot{\omega}_c$ and $D_c$ denote the reaction rate and the diffusivity of the progress variable in combustion DNS, respectively.

In case B with small $u'$, we observe that $S_d$ is always positive from both combustion DNS and propagating surfaces.
In cases D and E with larger $u'$, the variance of $S_d$ increases due to the increased turbulent fluctuations experienced by the reaction layer, and the probability of negative $S_d$ becomes higher, consistent with previous DNS results \citep[][]{Nivarti2017}.
The PDFs of $S_d$ from combustion DNS and propagating surfaces are qualitatively similar, which partly validates the model (\ref{eq:nonSd}).

\begin{figure}
  \centering
  \psfrag{a}[c][c]{\footnotesize $S_d/S_L$}
  \psfrag{b}[c][c]{\footnotesize PDF}
  \psfrag{i}{\footnotesize Case B}
  \psfrag{k}{\footnotesize Case D}
  \psfrag{l}{\footnotesize Case E}

  \subfigure{
  \centering
  \includegraphics[width=0.45\textwidth]{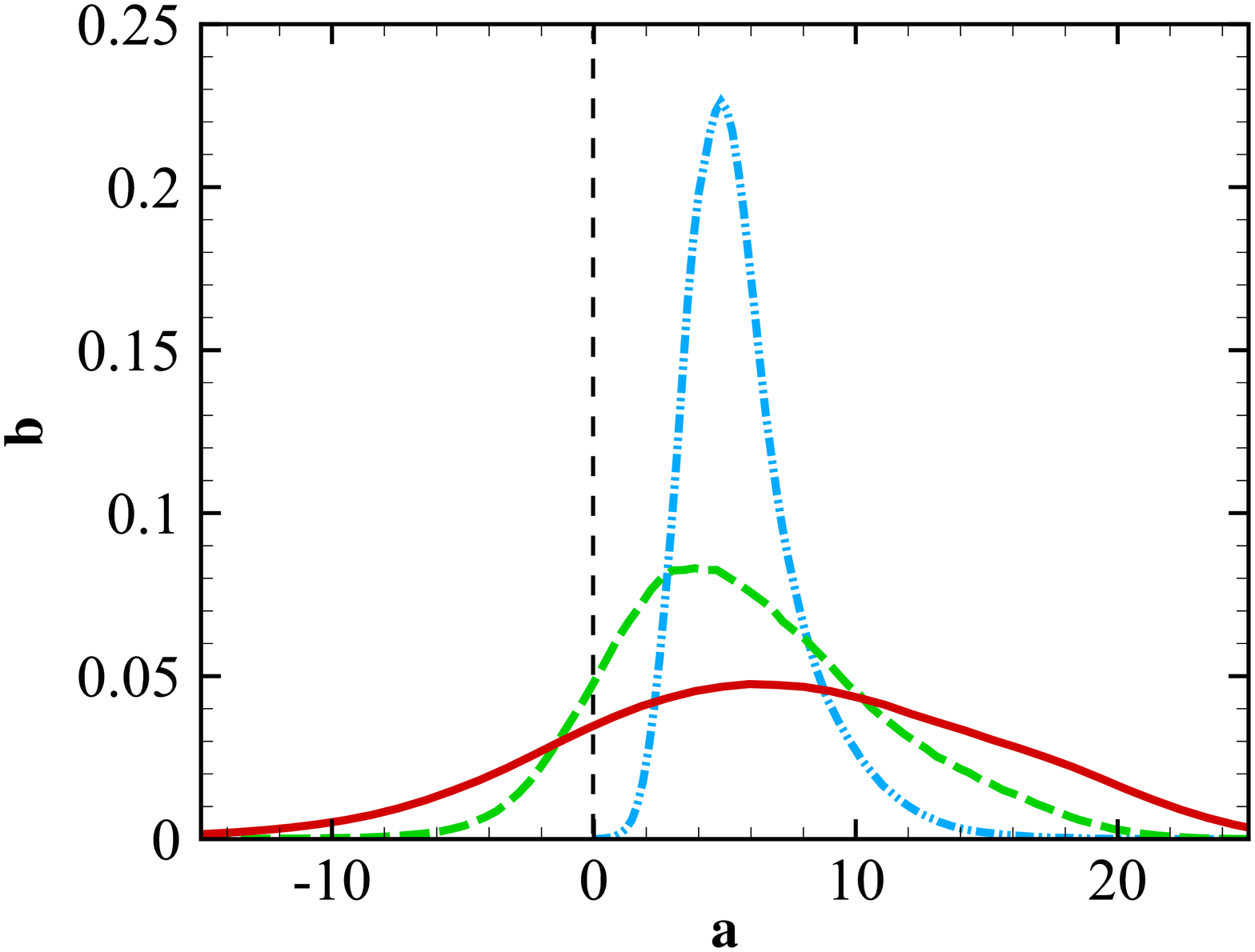}}
  \subfigure{
  \centering
  \includegraphics[width=0.45\textwidth]{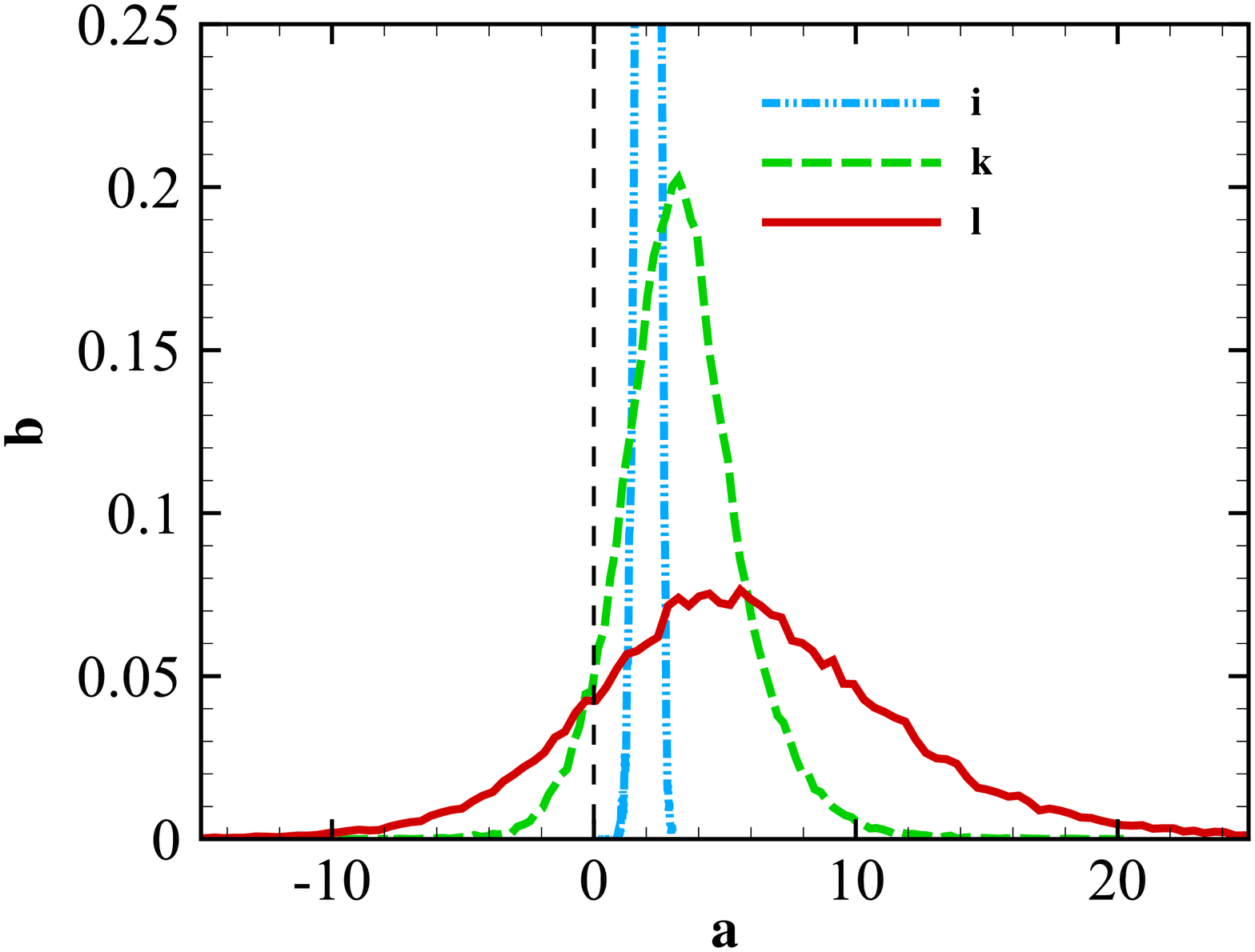}}
  \caption{(Colour online)~PDFs of the displacement speed for (a) flames in combustion DNS and (b) propagating surfaces in non-reacting DNS in cases B, D and E.}
  \label{fig:Sdpdf}
\end{figure}

\begin{figure}
  \psfrag{a}[c][c]{\footnotesize $t$~(ms)}
  \psfrag{b}[c][c]{\footnotesize $A(t)/A_0$}
  \centering
  \includegraphics[width=0.5\textwidth]{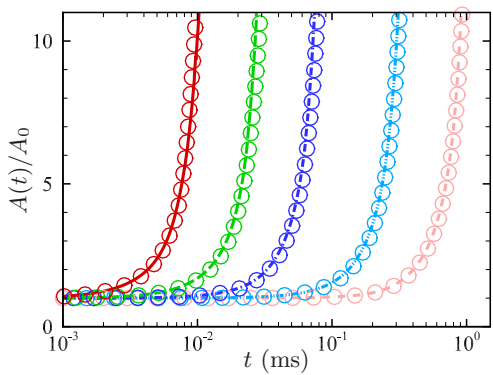}
  \caption{(Colour online)~The comparison of surface area ratios of propagating surfaces with constant $S_d$ (symbols) and non-constant $S_d$ (lines) in cases A--E (from right to left).}
  \label{fig:Sd_ATcomp}
\end{figure}

The evolutions of $A(t)/A_0$ of propagating surfaces with constant and non-constant $S_d$ are compared in figure~\ref{fig:Sd_ATcomp}. We find that the surface area ratios are almost identical, so the effect of non-constant $S_d$ of \eqref{eq:nonSd} on the prediction of $A_T/A_0$ using propagating surfaces appears to be negligible and we only use constant $S_d=S_L$ in the model development.

\section{Characteristic curvature of propagating surfaces}\label{sec:C_analysis}

The governing equation of $C^*$ can be derived from (\ref{eq:curve}) as \citep[see][]{Girimaji1992}
\EQ
\de{ C^*}{t^*}=u^*_{3,\alpha \beta} g_{\alpha \beta}-(K_t^*+2s^*_{\alpha \beta} g_{\beta \gamma}g_{\gamma \alpha})C^*+S^*_d g_{\alpha \beta} g_{\beta \gamma} g_{\gamma \alpha} {C^*}^2,
\label{eq:chf}
\EN
where $g_{\alpha \beta}=h_{\alpha \beta}/C^*$ is the normalized curvature tensor of unit norm. Taking the ensemble average over $N_s(t)$ surviving surface elements in (\ref{eq:chf}) yields
\EQ
\de{ \<C^*\>}{ t^*}=b_1+b_2 \<C^*\>+b_3 {\<C^*\>}^2,
\label{eq:chf_average}
\EN
where $b_1$, $b_2$, and $b_3$ denote the bending, straining and propagation terms, respectively.
The bending term initiates curvature on an initially planar surface, then straining and propagation effects take over.
The propagation term can cause a cylindrical surface with a finite initial curvature to develop a singularity.
Since $b_1$, $b_2$ and $b_3$ are functions of a statistically stationary Eulerian velocity $\bs u$ and a tensor $g_{\alpha\beta}$ of unit norm, they can be expected to be stationary random variables \citep[][]{Girimaji1992}.

We qualitatively analyze the exact solution to (\ref{eq:chf_average}) by assuming that $b_1, b_2$ and $b_3$ are constants.
Since in general the turbulent straining motion produces positive mean strain rate and increases the curvature of surfaces, statistically the stretching coefficient $b_2$ is negative from Eq.~(\ref{eq:chf}).
The surface elements are initially plane, so the initial condition for \eqref{eq:chf_average} is $\<C^*(0)\>=0$.

For material surface elements with $b_3=0$, we have
 \EQ
 \<C^*(t^*)\>=\frac{b_1}{b_2}(\exp(b_2 t^*)-1),
 \EN
with $\lim_{t^*\rightarrow\infty}\<C^*(t^*)\> = b_1/|b_2|$.
For propagating surfaces, we discuss solutions based on the discriminant $b_2^2 - 4b_1b_3$.
If $b_2^2 - 4b_1b_3=0$, then we have
 \EQ
 \<C^*(t^*)\>= \frac{b_2^2 t^*}{2b_3(2-b_2t^*)}
\EN
with  $\lim_{t^*\rightarrow\infty}\<C^*(t^*)\> = |b_2|/(2b_3)$; if $b_2^2 - 4b_1b_3 \equiv {\Delta_I}^2>0$, then we have
 \EQ
 \<C^*(t^*)\>=2b_1\frac{\exp({\Delta_I} t^*)-1}{b_2+{\Delta_I} -(b_2- {\Delta_I} )\exp({\Delta_I}t^*)},
 \EN
with $\lim_{t^*\rightarrow\infty}\<C^*(t^*)\> = 2b_1/({\Delta_I}-b_2)$; if $4b_1b_3-b_2^2 \equiv {\Delta_{II}}^2>0$, then we have
  \EQ
 \<C^*(t^*)\>=\frac{-b_2+\Delta_{II} \tan{ \left[ \frac{\Delta_{II}}{2}t^*-\arctan \left(\frac{b_2}{\Delta_{II}}\right) \right] }}{b_3},
 \EN
where $\<C^*(t^*)\>$ is periodic in time but always finite.

This qualitative analysis implies that the solution to \eqref{eq:chf_average} is finite as $t^*\rightarrow\infty$, which supports the occurrence of the stationary state of $\<C^*\>$ shown in figure~\ref{fig:chf}.

\section{Comparison of model predictions of $S_T$}\label{sec:model_comp}
\begin{figure}
  \centering
  \psfrag{x}{\footnotesize $u'/S_L$}
  \psfrag{y}{\footnotesize $S_T/S_L$}
  \psfrag{a}{\footnotesize{DNS}}
  \psfrag{b}{\footnotesize{Present model \eqref{eq:STmodel_predict}}}%
  \psfrag{c}{\footnotesize{Model \eqref{eq:ST_linear}}}
  \psfrag{d}{\footnotesize{Model \eqref{eq:STmodel_Klimov}}}
  \psfrag{e}{\footnotesize{Model \eqref{eq:STmodel_Zimont}}}
  \psfrag{f}{\footnotesize{Model \eqref{eq:STmodel_Bradley}}}
  \psfrag{g}{\footnotesize{Model \eqref{eq:STmodel_Kawanabe}}}
  \psfrag{h}{\footnotesize{Model \eqref{eq:STmodel_Peters}}}
  \includegraphics[width=0.7\textwidth]{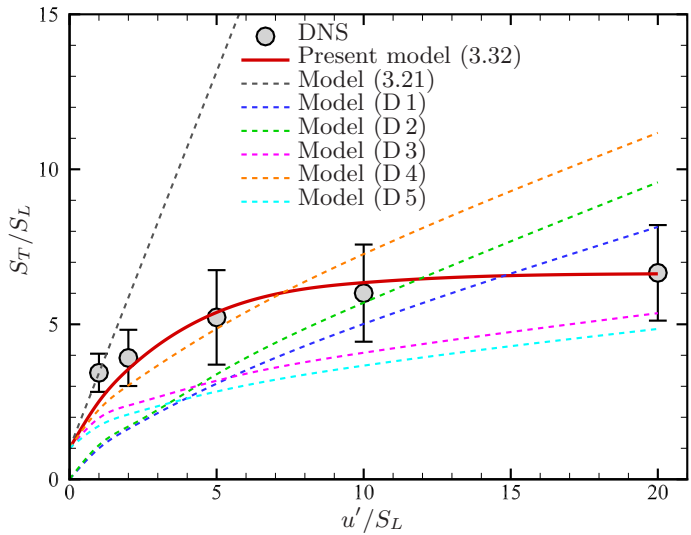}

  \caption{(Colour online)~Comparison of $S_T$ calculated from the present combustion DNS, the present model and  various existing models. The error bar denotes the standard deviation of $S_T/S_L$ in the DNS.}
  \label{fig:STmodel_comparison}
\end{figure}

We compare the model predictions of $S_T$ among the present model and some existing models listed below.
\begin{enumerate}[(1)]
\setlength{\labelsep}{0.3em}
  \item Linear model \eqref{eq:ST_linear} with $\mc C$ listed in table~\ref{tab:parampredict}.
  \item Model suggested by \citet[]{Klimov1983}:
  \EQ
  S_T=C_K u'^{0.7}S_L^{0.3}~\textrm{with}~C_K=1.
  \label{eq:STmodel_Klimov}
  \EN
  \item Model suggested by \citet[]{Zimont1988}:
  \EQ
  S_T = C_Z u' Da^{1/4}~\textrm{with}~C_Z = 1.
  \label{eq:STmodel_Zimont}
  \EN
  \item Model suggested by \citet[]{Bradley1992}: 
  \EQ
  \frac{S_T}{S_L}=1+0.95Le^{-1}\left(\frac{u'}{S_L}\frac{l_t}{\delta_L}\right)^{1/2},~\textrm{with}~Le=1.
  \label{eq:STmodel_Bradley}
  \EN
  \item Model suggested by \citet[]{Kawanabe1998}:
  \EQ
  \frac{S_T}{S_L}=1+1.25\left(\frac{u'}{S_L}\right)^{0.7}.
  \label{eq:STmodel_Kawanabe}
  \EN
  \item Model suggested by \citet[]{Peters1999}: 
  \EQ
  \frac{S_T}{S_L}=1+\frac{0.39}{2}\frac{l_t}{\delta_L}(\sqrt{1+20.5/Da}-1).
  \label{eq:STmodel_Peters}
  \EN

\end{enumerate}
Most of these models involve arbitrarily adjustable or empirical constants, and these model constants are tuned to give the best prediction of $S_T$ for the present combustion DNS. The comparison is provided in figure~\ref{fig:STmodel_comparison}.

We observe that the linear model is only valid for very small $u'$.
All the other existing models can predict the trend of bending at moderate $u'$, but they have significant discrepancies from the DNS result, either over- or under-predictions, because these models may only be valid at certain flame/turbulence parameters and flame geometry/boundary conditions.
We remark that the variation of instantaneous $S_T$ is large in the present combustion DNS from error bars in figure~\ref{fig:STmodel_comparison}, though the mean $S_T$ from the present DNS has been averaged over 20 large eddy-turnover times.
Previous DNS studies of turbulent premixed flames \citep[\eg][]{Aspden2011,Savard2015b} also observed the notable uncertainty of $S_T$ owing to the unstable flame and the finite size of computational domain.
Thus small discrepancies between the model prediction and DNS result should be acceptable owing to the uncertainties in the DNS result.

\begin{figure}
  \centering
  \psfrag{x}{\footnotesize $u'/S_L$}
  \psfrag{y}{\footnotesize $\epsilon_{\textrm{model}}$}
  \psfrag{a}{\footnotesize{Present DNS}}
  \psfrag{b}{\footnotesize{\cite{Nivarti2017}}}
  \psfrag{c}{\footnotesize{\cite{Lee2010}}}

  \includegraphics[width=0.7\textwidth]{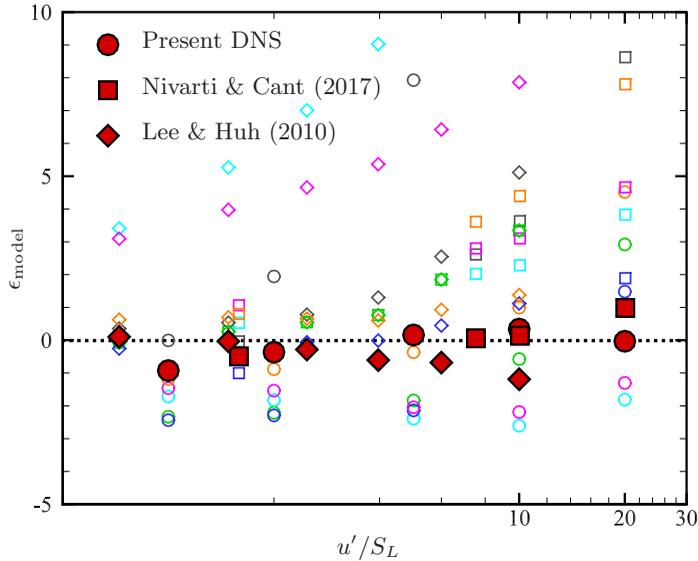}
  \caption{(Colour online)~Comparison of the discrepancies from various model predictions of $S_T$ in three DNS series.
  Symbols denote different DNS series. Circles: the present DNS, diamonds: \cite{Nivarti2017}, squares: \cite{Lee2010}.
  Colours denote different models. Red: the present model \eqref{eq:STmodel_predict}, gray: model \eqref{eq:ST_linear}, blue: model \eqref{eq:STmodel_Klimov}, green: model \eqref{eq:STmodel_Zimont}, magenta: model \eqref{eq:STmodel_Bradley}, orange: model \eqref{eq:STmodel_Kawanabe}, light blue: model \eqref{eq:STmodel_Peters}.
  The dotted line denotes $\epsilon_{\textrm{model}}=0$.}
  \label{fig:STmodel_comparison_scatter_abs}
\end{figure}

To further compare model performances in all the three DNS series, we define the model discrepancy as
\EQ
\epsilon_{\textrm{model}}=\frac{S_T^{\textrm{model}}-S_T^{\textrm{DNS}}}{S_L},
\label{eq:deviation}
\EN
where $S_T^{\textrm{model}}$ and $S_T^{\textrm{DNS}}$ are obtained from models and DNS, respectively.
In figure~\ref{fig:STmodel_comparison_scatter_abs}, the present model provides consistently good predictions for a wide range of $u'$ and various fuels, whereas other model predictions are very scattered with notable model discrepancies.


\bibliographystyle{jfm}
\bibliography{JFM_ST_PS}

\end{document}